\documentclass[preprint,12pt]{elsarticle}

\usepackage{hyperref}
\usepackage{amssymb}
\usepackage{graphicx}
\usepackage{subcaption}
\usepackage{algorithm2e}
\usepackage{algorithmic}
\usepackage{booktabs,caption}
\usepackage[flushleft]{threeparttable}

\usepackage[table,xcdraw]{xcolor}
\usepackage{fourier} 
\usepackage{array}
\usepackage{makecell}
\usepackage{hyphenat}

\journal{arXiv}

\begin{document}

\begin{frontmatter}



\title{From Malware Samples to Fractal Images: A New Paradigm for Classification\\Version 2.0, \\Former paper name: Have you ever seen malware?}


\author[inst2]{\href{https://orcid.org/0000-0002-3858-7340}{\texorpdfstring{\includegraphics[scale=0.06]{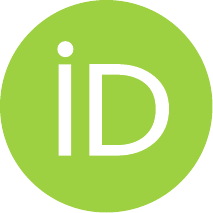}\hspace{1mm}Ivan Zelinka}}}


\author[inst2]{Miloslav Szczypka}
\author[inst2]{Jan Plucar}
\author[inst3,inst4]{Nikolay Kuznetsov}

\affiliation[inst2]{organization=
            Department of Computer Science, FEI
			VSB Technical University of Ostrava
			Tr. 17. Listopadu 15, Ostrava
			Czech Republic
			Email: {ivan.zelinka, miloslav.szczypka, jan.plucar}@vsb.cz,
			www.ivanzelinka.eu
			}
			
\affiliation[inst3]{organization=
	Institute for Problems in Mechanical Engineering RAS, 199178 St. Petersburg, Russia
	Email: nkuznetsov239@gmail.com}

\affiliation[inst4]{organization=
	Faculty of Mathematics and Mechanics, St. Petersburg State University, 198504 Peterhof, St. Petersburg, Russia
	Email: nkuznetsov239@gmail.com
}

\begin{abstract}
To date, a large number of research papers have been written on the classification of malware, its identification, classification into different families and the distinction between malware and goodware. These works have been based on captured malware samples and have attempted to analyse malware and goodware using various techniques, including techniques from the field of artificial intelligence. For example, neural networks have played a significant role in these classification methods. Some of this work also deals with analysing malware using its visualisation. These works usually convert malware samples capturing the structure of malware into image structures, which are then the object of image processing. In this paper, we propose a very unconventional and novel approach to malware visualisation based on dynamic behaviour analysis, with the idea that the images, which are visually very interesting, are then used to classify malware concerning goodware. Our approach opens an extensive topic for future discussion and provides many new directions for research in malware analysis and classification, as discussed in conclusion. The results of the presented experiments are based on a database of 6 589 997 goodware, 827 853 potentially unwanted applications and 4 174 203 malware samples provided by ESET\footnote{https://www.eset.com/} and selected experimental data (images, generating polynomial formulas and software generating images) are available on GitHub\footnote{TBA if accepted} for interested readers. Thus, this paper is not a comprehensive compact study that reports the results obtained from comparative experiments but rather attempts to show a new direction in the field of visualisation with possible applications in malware analysis.
\end{abstract}




\end{frontmatter}


\section{Introduction}
\label{Introduction}
The issue of malware, essentially the alpha and omega of cybersecurity, has become very important in recent years. There are now many methods for identifying and classifying malware, either into malware families or distinguishing it from goodware.
Most of the research during the past years was focused on static (traditional analysis, image analysis), dynamic (analysis of behaviour) and other analysis methods. In static analysis, the structure information of binary executable files is sequenced, and the behaviour of this code is analysed.
The methods can be specified as follows:

\begin{itemize}
\item Static method with static features - extracts opcodes, strings and sequences of bytes,  analysing them, with a possible extraction of structural information of a binary code. 
\item Static method with visual analysis - grayscale or RGB images using one or multi-dimensions to analyse inherent features extracted from data samples.
\item Dynamic method/behaviour graph method  - analysing the behaviour of the running program to determine whether it is malware, as for example, by building a behaviour graph for each known malware family.
\end{itemize}

Our research is focused on analysing malware detection and classification by a very specific graphical method that extends contemporary state-of-the-art. Let's take a closer look on above mentioned methods. 

\subsection{Static method with visual analysis}

The analysis of malware using image processing can be given in two ways, namely static analysis and dynamic analysis. The dynamic analysis consists of recording the behaviour of the malware in the protected environment and then converting it to an image, which is usually grayscale and resembles the grain on older TV screens. Static analysis is based on working with the binary file and converting it into a similar images in a various ways. 
These approaches can be seen in publications \cite{ANOVEL2021} discussing deep learning in malware image classification, \cite{MALWARECLASS2020} describing malware classification using linear transformation, \cite{AMULTICHANNEL2019} showing multi-channel visualization method for malware classification based on deep learning, \cite{MALWAREIDENT2018} using malware classification algorithm that uses a static function called MCSC (Malware Classification using SimHash and CNN), converting malware codes into grey-scale images based on SimHash and then identifies their families using convolutional neural network algorithms, \cite{DETECTINGANDROIDMALWARE2017} discussing the inspected android APK files, using malware images patterns, and \cite{MACHINELEARNANDROID2016} using machine learning based method detecting android malware by analyzing the visual representation of binary formatted APK file into Grayscale, RGB, CMYK and HSL. 

A novel deep neural network called SERLA is reported in \cite{ANOVEL2021}. In the solution proposed in this research \cite{ANOVEL2021}, due to the influence of the used method, the speed of image generation is slow without using multithreading. The complexity of the SERLA network significantly increases the model training time.

Both types of files - bytes files (binary byte data of malware and clean files), together with asm files (disassembly file of malware and clean files) were processed into graphical images representing the structure of data, then analyzed and the decision made if they are malware. The representation method was based on the binaries, where word vectors were extracted from both bytes files and asm files. This is similar to the research \cite{AMULTICHANNEL2019}, where the CBOW model of Word2Vec was used to calculate word vectors for each assembly instruction.

A much simpler solution \cite{MALWARECLASS2020}, offered by a team of scientists, proposing a deep neural network called MCSC (Simhash combined with CNN), where 3 phases were used to process malware data. In this case, the team compared their training model with FC Classifier, tanhMCSC, reluMCSC, MCP, and MCSLT, outperforming them based on experimentational results. 

One interesting research has been performed using the traditional static detection method, utilizing the visual analysis method, but without any deep neural network, \cite{MACHINELEARNANDROID2016}. The authors collected data from various sources (malware dataset and goodware apps dataset), and then converted them into all four representations: Grey, RGB, CMYK and HSL, then extracted features using GIST (summarizes gradient information (scale, orientation of image parts using Gabor filter)), and then trained three machine learning (ML) algorithms (Decision Tree,  Random Forest (RF) and kNearest Neighbor (kNN)) with this data as input, after which the end performance of each ML was evaluated, and the best method selected.
This method used APK files for samples and utilized a multi-channel visualization technique. The files were unpacked and processed into graphical images representing the structure of data, then analyzed, and the decision was made if they were malware or not.

Another interesting method is also based on a machine learning algorithm, which is not using any deep neural network architectures \cite{DETECTINGANDROIDMALWARE2017}. This research uses an image-based classification (Random Forest classifier) and texture features to characterize and classify malware. This way, a number of decision trees are constructed at training time, and a malware class is produced at the output. A malware dataset (Android Drebin\footnote{https://www.sec.tu-bs.de/~danarp/drebin/}) was used together with a clean apps dataset (Baidu Apps Market) for training and testing purposes.  

\subsection{Dynamic method}

The dynamic analysis method has generally higher accuracy than the first two methods but is more time-consuming. It executes malware samples, usually in a simulated safe environment, to monitor system calls via processing APIs and automatically generate detailed reports.

A dynamic detection method \cite{MALWAREBEHAVIORIMAGE2014} was used with machine learning algorithms for the detection and classification of malware. A behaviour-based technique was used for visualizing malware behaviour in the form of images. 
Researchers claim that it is possible to achieve high accuracy (up to 99.33\%) of malware classification in identifying variants using malware behaviour images.

Another method solves the problem of slow analysis methods mentioned above and suggests focusing on known malware families and not on individual malware cases in order to simplify and speed up the malware analysis and detection process. 
Many solutions, as already mentioned \cite{ANOVEL2021} \cite{DETECTINGANDROIDMALWARE2017} \cite{MACHINELEARNANDROID2016} and \cite{MALWAREBEHAVIORIMAGE2014}, fall into this category as well.

\subsection{Image Binary Data Analysis}

One more area where the graphical algorithms can be used for malware detection and classification is steganography \cite{INTENSIVE}. The paper discusses various techniques for embedding malicious payloads in images and proposes its own solution to detect these methods. This is the area where antiviruses and antimalware programs fail while relying on traditional techniques, such as specific detection, generic detection, and heuristic detection. 

In work, \cite{INTENSIVE}, the team of researches checked graphical images with Aho-Corasick substring search algorithm, where in EXIF data of the image file, they were looking for certain PHP keywords in image files. In case such strings were found, this usually indicated that these files are contaminated with some kind of hidden information, which may be part of some malware family, thus, have a harmful nature.

In \cite{venkatraman2019hybrid}, the authors propose a unified hybrid approach for malware identification based on deep learning and visualization for effective malware detection. They have two objectives, namely to present the use of new techniques to detect suspicious system behaviour and to explore the use of hybrid image-based approaches along with deep learning architecture for more effective malware classification. They measure the performance of their method using measures of similarity of malware behaviour patterns and also using a configuration-sensitive deep learning architecture.

In the paper \cite{kancherla2013image}, the authors go deeper and outline the issue of metamorphism, obfuscation and other stealth technologies used by modern malware. They also present a visualization-based approach to malware detection, with executables converted to grayscale images. From these, low-level features related to intensity and texture are then extracted, and then the authors use intelligent algorithms to detect malware. This is based on a relatively large sample size of 25 000 malware samples and 12 000 goodware samples.

To further improve the quality of identification, the authors of \cite{vasan2020image} propose a new CNN-based architecture that is said to be effective for detecting packed and unpacked malware. According to the authors, CNN provides different semantic representations of images, and a set of these architectures can extract features with higher quality than traditional methods. Their method is particularly suitable for malware detection and reports a relatively high accuracy of around 98\%.

Virtually all methods that work with images are, with rare exceptions, methods that work with grayscale granular images. The transformations used to create these images have not accounted for the impact of the encoding and the technique of rendering pixels in colour in terms of machine learning classifier performance. This is addressed by the authors of the work \cite{vu2020hit4mal}, who propose a new approach to encoding and arranging the bytes from binaries into images so that they contain statistical and syntactic artefacts and are in colours. According to the authors, the accuracy of their method is 93\%.

\subsection{Fractal Geometry and Malware Analysis}
Apart from classical methods based on static or dynamic analysis, either of code or its visual representation, research in the field of malware classification has touched upon such exotic areas as fractal geometry which is also part of our paper. These papers \cite{jaenisch2012fractals}, \cite{nisa2020hybrid} approach malware identification in a different way than our paper and address the following.

In \cite{jaenisch2012fractals} the authors put forward the hypothesis that the boundary between malware and goodware is of a fractal nature. Based on this hypothesis, they then introduce a coding method derived from disassembling programs and converting them to optcode. These optcodes are then converted into real numbers and used to characterize the frequency of malware functions compared to the properties of goodware functions. From this identified information, they then derive classifiers based on which they interpolate and extrapolate the parameter space of the samples to identify the boundaries of the parameter space. They report that preliminary results strongly support the hypothesis of a fractal boundary between goodware and malware.

As the number of malware and its families grows, there is naturally a growing demand for classification methods that can detect pathological codes. This is the topic addressed by the researchers in the \cite{nisa2020hybrid} paper, where they propose a symptom fusion method that combines symptoms from retraining deep neural networks such as AlexNet and Inception-V3. It combine this with the information obtained by segmentation fractal texture analysis of images that represent malicious code. This again works with greyscale images. To improve the classification, the authors also use the so-called affine transformations \cite{barnsley2014fractals} of the  image. Affine transformations are a fundamental part and basis of fractal geometry. In that paper, the authors classify malware into 25 families.

In the area of fractal geometry applied to cybersecurity, there are of course, other publications such as \cite{khan2017cognitive}, \cite{cowen2013fractal}, \cite{khan2018malvidence} or \cite{ren2020malware}, which also try to apply fractal geometry to the classification of malware into different families or differentiation from goodware. It must be said, however, that there are \textbf{multiple} fewer of these publications than publications that use classical methods and representations. Authors look at the malware problem through the lens of fractal geometry, mostly trying to address fractal dimensions, fractal boundaries between malware and goodware domains, and so on. Some of the aforementioned approaches and methods are outlined in Tab. \ref{methodsoverview}.

\begin{table}[!ht]
    \centering
    \tiny
        \caption{Malware identification, various methods.}
    \begin{tabular}{|l|l|l|l|l|l|}
    \hline
        Source & Year & Data Analysis & Dataset & Classification Approach & Accuracy in \% \\ \hline 
        \hline
        \cite{narayanan2018multi} & 2018 & Static & Mw, Gw & Graph Kernels, SVM & 94.00 \\ \hline
        \cite{du2019novel} & 2018 & Dynamic & Mw & SVM, Decision Tree, Naïve Bayes & 88.30 \\ \hline
        \cite{alam2017droidnative} & 2018 & Static & Mw, Gw & SVM & 93.22 \\ \hline
        \cite{kang2015detecting} & 2017 & Static & Mw, Gw & SVM & 90.00 \\ \hline
        \cite{wen2017android} & 2018 & Hybrid & Mw, Gw & SVM & 95.20 \\ \hline
        \cite{cui2018detection} & 2018 & Hybrid & Mw & SVM & 94.50 \\ \hline
        \cite{vasan2020image} & 2020 & Static & Mw & Ensemble of convolutional neural networks & 99.50 \\ \hline
        \cite{nataraj2011malware} & 2011 & Static & Mw & Nearest Neighbor & 97.18 \\ \hline
        \cite{anderson2012improving} & 2012 & Static, dynamic & Mw & SVM & 98.00 \\ \hline
        \cite{dahl2013large} & 2013 & Dynamic & Mw, Gw & Neural networks and logistic regression & 86.00 \\ \hline
        \cite{zhang2014semantics} & 2014 & Dynamic & Mw & Semantics-based & 93.00 \\ \hline
        \cite{pascanu2015malware} & 2015 & Static & Mw, Gw & Logistic regression and MLP classifier & 98.30 \\ \hline
        \cite{garcia2016random} & 2016 & Static & Mw & Random Forest & 95.00 \\ \hline
        \cite{moshiri2017malware} & 2017 & Dynamic & Mw & Machine learning techniques & 99.00 \\ \hline
        \cite{liu2017automatic} & 2017 & Static & Mw, Gw & SNN clustering algorithm & 98.90 \\ \hline
        \cite{cakir2018malware} & 2018 & Static & Mw & Gradient boosting & 96.00 \\ \hline
        \cite{kalash2018malware} & 2018 & Static & Mw & CNN-based architecture & 98.50 \\ \hline
        \cite{naeem2019identification} & 2019 & Static & Mw & SVM, KNN & 98.00 \\ \hline
        \cite{cui2018detection} & 2019 & Static & Mw & CNN & 97.60 \\ \hline
        \cite{naeem2019detection} & 2019 & Static & Mw & DCNN & 98.18 \\ \hline
        \cite{ren2020malware} & 2019 & Static & Mw & Space filling (fractal) curve mapping & \thead{Multiple experiments\\ 98.36\% and 99.08\% \\99.21\% and 98.74\%} \\ \hline
        \cite{khan2018malvidence} & 2018 & Static & Mw & Fractal geometry, graph theory & \thead{Various experiments,\\thesis, max 96} \\ \hline
        \cite{cowen2013fractal} & 2013 & Static & Mw & Space filling (fractal) curves, random walk & \thead{Noncomparative study\\ Graphical method of\\polymorphism identification}  \\ \hline
        \cite{MALWAREBEHAVIORIMAGE2014} & 2014 & Dynamic & Mw & API sequences mapped into heat maps. & 99.33 \\ \hline
        \cite{vu2020hit4mal} & 2019 & Static & Mw, Gw &  Hybrid image transformation & 93.01 \\ \hline
        \cite{vasan2020image} & 2020 & Static & Mw & Ensemble of CNNs & 99.00 \\ \hline
        \cite{kancherla2013image} & 2013 & Static & Mw & SVM & 95.00 \\ \hline
        \cite{venkatraman2019hybrid} & 2019 & Static & Mw & Deep learning & \thead{Multiple experiments\\ 89.3 - 91.6} \\ \hline
        \cite{khan2017cognitive} & 2017 & Static & Mw & Graph theory, fractal geometry & Noncomparative study \\ \hline
        \cite{nisa2020hybrid} & 2020 & Static & Mw & Fractal texture analysis, deep learning & 99.30 \\ \hline
    \end{tabular}
    \label{methodsoverview}
\end{table}

Our approach is both the same and, at the same time, very different. Same in that we also use fractal geometry and different in that we use it to visualize individuals of malware and goodware samples as the fractal objects. Then deep learning is used to classify it.

\subsection{What next?}
From the above albeit very brief overview, which cannot cover all the important research works due to the space in the paper, it can be seen that malware research in the sense of identification is not only focused on identification using classical methods, but recently, there is an increasing number of works dealing with malware visualization by analyzing the attributes of such images and classifying malware into families or distinguishing it from goodware. We can see works that convert binaries directly into images, as well as works that perform disassembly using reverse engineering and convert the code thus obtained back into a grayscale image after modifications. It can also be observed that the performance of these methods is more or less between 90\% to 99\%, exceptionally below 90\%. 

It can also be observed that the success of these methods certainly depends on which classification algorithms were used, in the case of deep learning, how large the network was, what its structure was and how many learning parameters it contained. Furthermore, how balanced the dataset was, is important as some papers report a diametric difference in the malware and goodware samples used. More specifically, the malware dataset is relatively often significantly larger than the goodware dataset, and the number of samples tends to be on the order of thousands to tens of thousands. However, it is possible to encounter works with hundreds of samples. In this case, however, these are usually works that classify malware into different families. From these findings, it is clear that more or less every method has a chance for further improvement and that the problem of processing malware visualization is not yet fully closed and is just beginning its dawn, all the more so because the AI methods that are involved in this are beginning to strengthen in their capabilities significantly. 

This is where our work fits in, proposing a radically different approach to malware visualization along with basic classification experiments and attempting to bring this topic to the expert community for further discussion and development.

\section{Motivation, the paper contribution and the structure}
Based on the analysis of the current state of the art, it is pretty clear that the issue of malware identification is not only very complex but also that it is a very topical issue. When studying the current state of the art in malware identification and classification research in the form of visual representation of malware structure, it is clear that not all possibilities of graphical representation have been exhausted. One possibility is the so-called fractal geometry. Its first signs were discovered at the end of the 19th century, and it was officially mathematically formalized and published by the famous mathematician B.B. Mandelbrot in  \cite{mandelbrot1977fractal}. Fractals have a remarkable peculiarity in that they can visually and mathematically capture - demonstrate extremely complex objects. It has also been proven that there are an infinite number of fractals, \cite{barnsley2014fractals}, \cite{mandelbrot1977fractal}. The question is whether we can find suitable representations that, based on analysis of the structure or behaviour of malware, can represent such malware as a fractal. This can then be subsequently analyzed by a suitable image-processing method. The advantage of such a representation is that infinitely many fractal sets can be generated. It is thus a reservoir with an infinite number of patterns that can also be assigned to future malware according to appropriate criteria. In this paper, we demonstrate some of the results of our research, which show that this direction is very promising and that much more research work can be done in this direction. Thus, this paper is not a comprehensive study that both opens and closes the topic but rather an example of a new area that other researchers can consider.

\begin{figure}[!ht]
    \centering
    \includegraphics[scale=0.4]{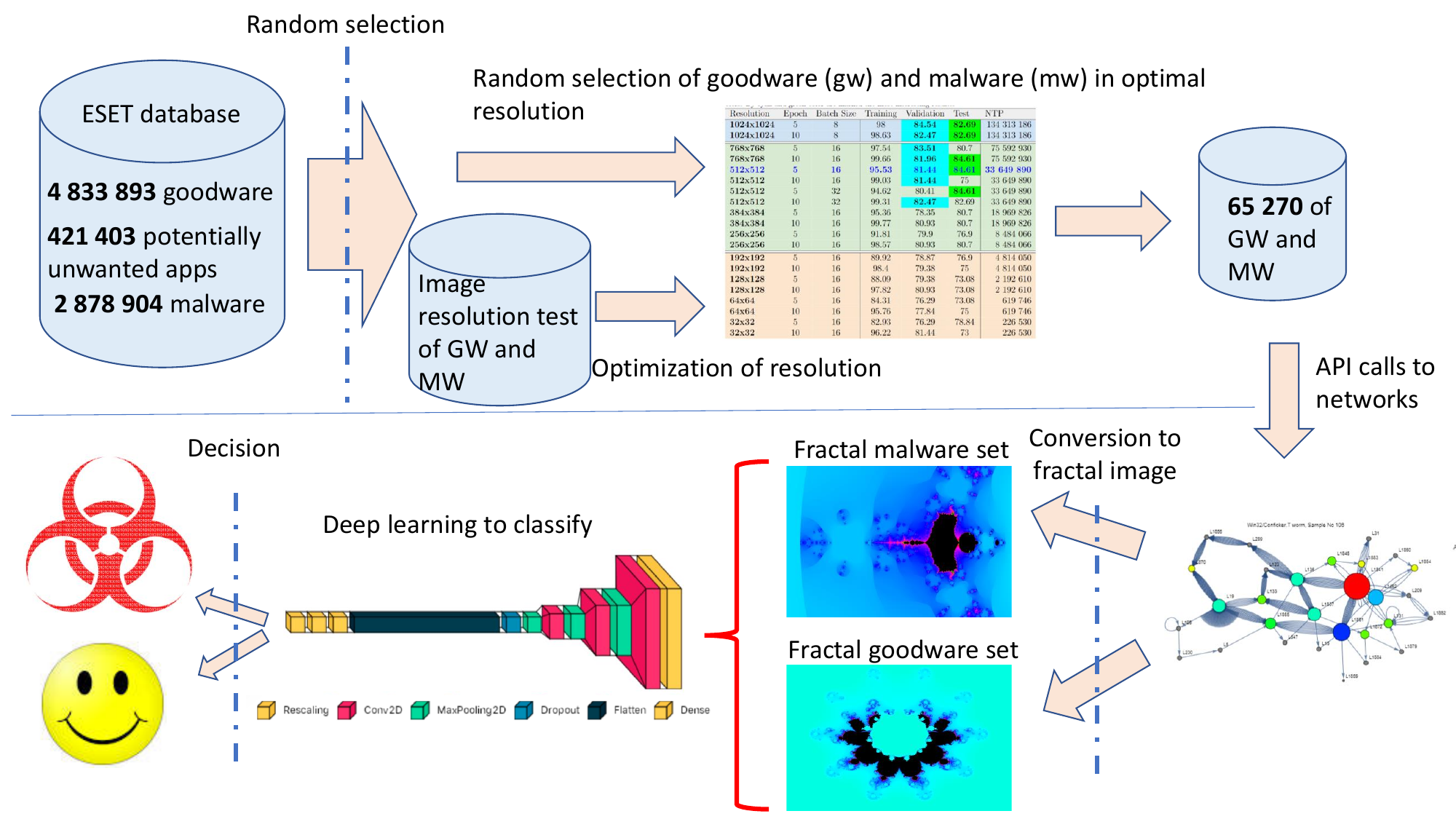}
    \caption{A flowchart of our experiments.}
    \label{flow}
\end{figure}

The structure of this paper is as follows. After a brief mention of fractal geometry, we focus on a discussion of our experiments, in particular, the data used. We then go on to discuss how this data was converted into its final form and how it was used to render the fractal patterns that more or less characterize the malware used. This is then followed by a section that deals with experiments in image processing using deep learning, where we show how such patterns could be recognized and thus classify goodware and malware. The principle outline of how we performed the experiments in this paper is shown in Figure \ref{flow}.

\begin{figure}[!ht]
\begin{minipage}[b]{0.5\linewidth}
\centering
\includegraphics[scale=0.05]{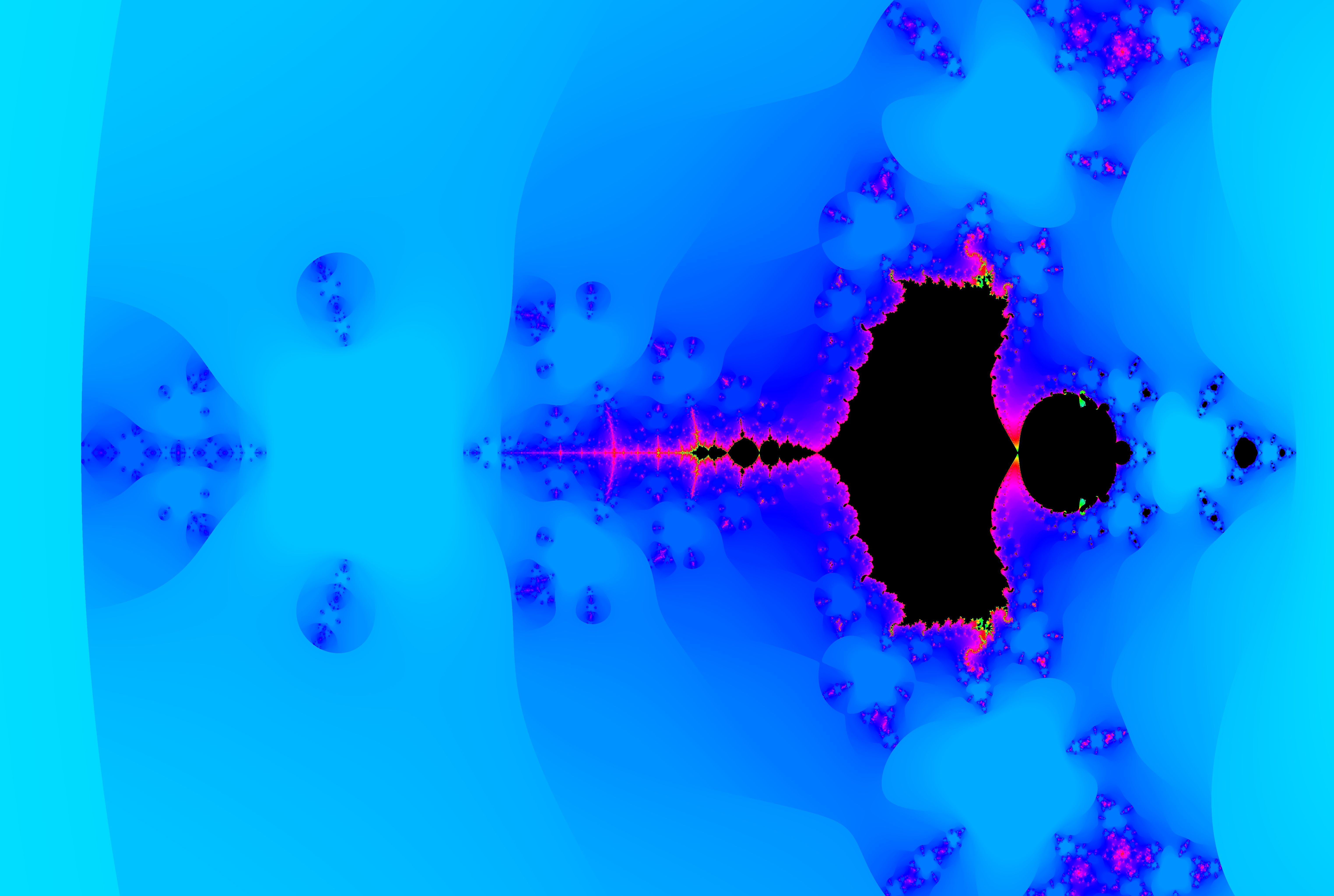}
\caption{Win32.Doomer.E worm \\ alias The Batman.}
\label{batman}
\end{minipage}
\hspace{0.5cm}
\begin{minipage}[b]{0.5\linewidth}
\centering
\includegraphics[scale=0.05]{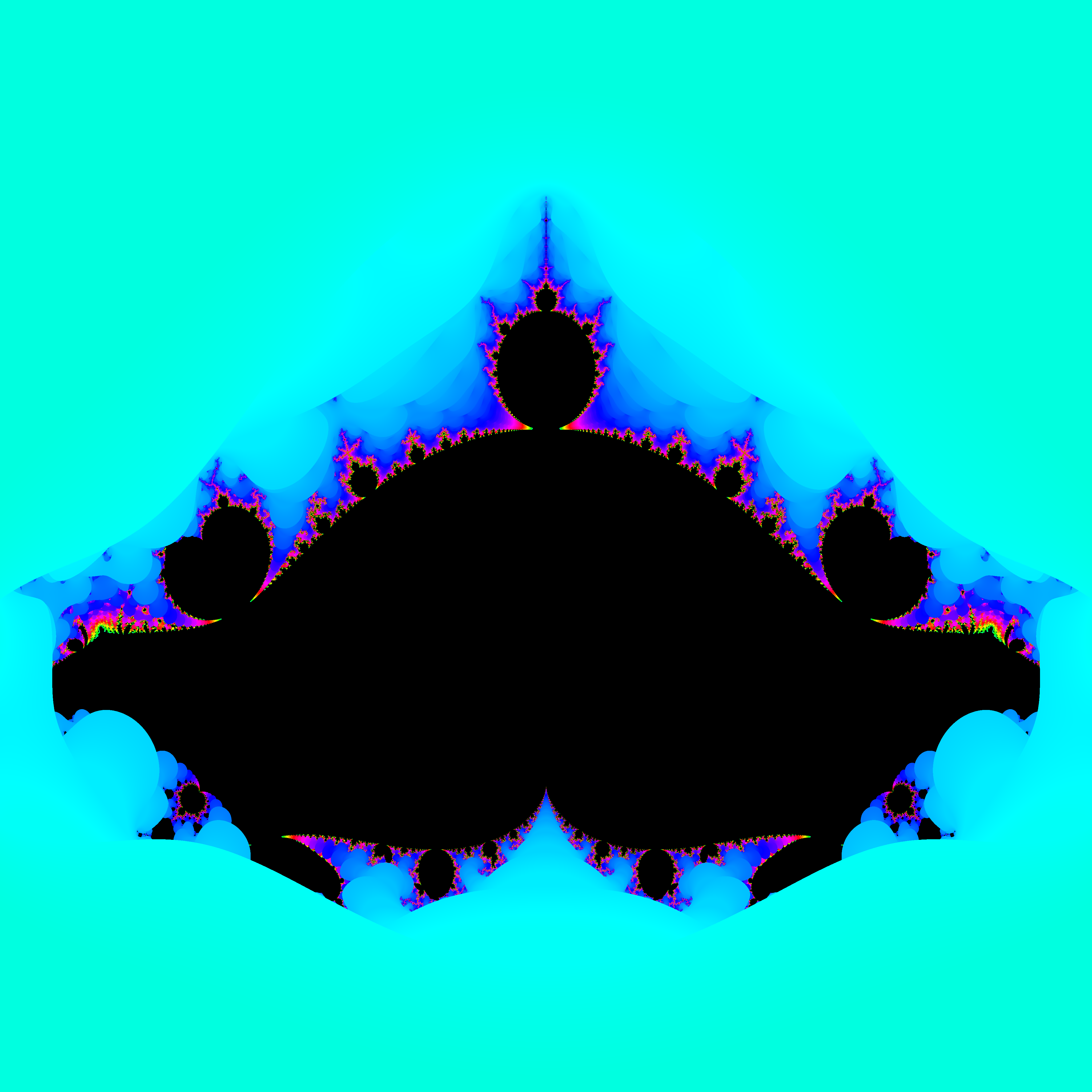}
\caption{Win32.Conficker.AQ worm \\ alias the Temple of Buddha.}
\label{budha}
\end{minipage}
\end{figure}

In the end, we also raise a number of research questions that we consider very important. Answering them could provide further interesting information in the field of malware research through the lens of fractal geometry. Another and, in our opinion, not insignificant contribution, which is unfortunately very much overlooked today, is the aesthetic benefit that comes from the use of fractal geometry. The latter shows a very charming (cf. Fig. \ref{batman}, \ref{budha}, which we have given nicknames to, or gallery in section \ref{gallery}) combination of science and computer art, which is unfortunately not seen much in science today. Therefore, the results of our research can be used not only to further new directions in computer malware research but also in the attractive popularization of such research for the non-expert public.

\section{Fractal geometry} 
\begin{figure}[!ht]
\begin{minipage}[b]{0.5\linewidth}
\centering
\includegraphics[scale=0.8]{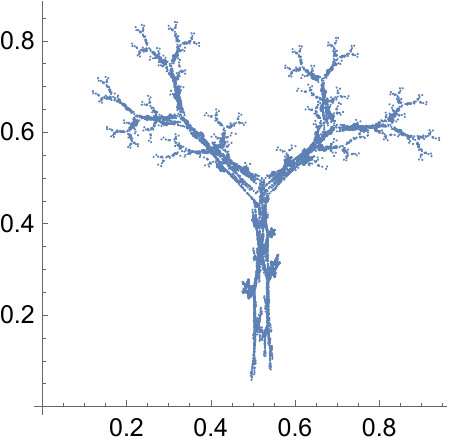}
\caption{Fractal \textit{Tree} plotted by IFS algorithm.}
\label{tree}
\end{minipage}
\hspace{0.5cm}
\begin{minipage}[b]{0.5\linewidth}
\centering
\includegraphics[scale=0.4]{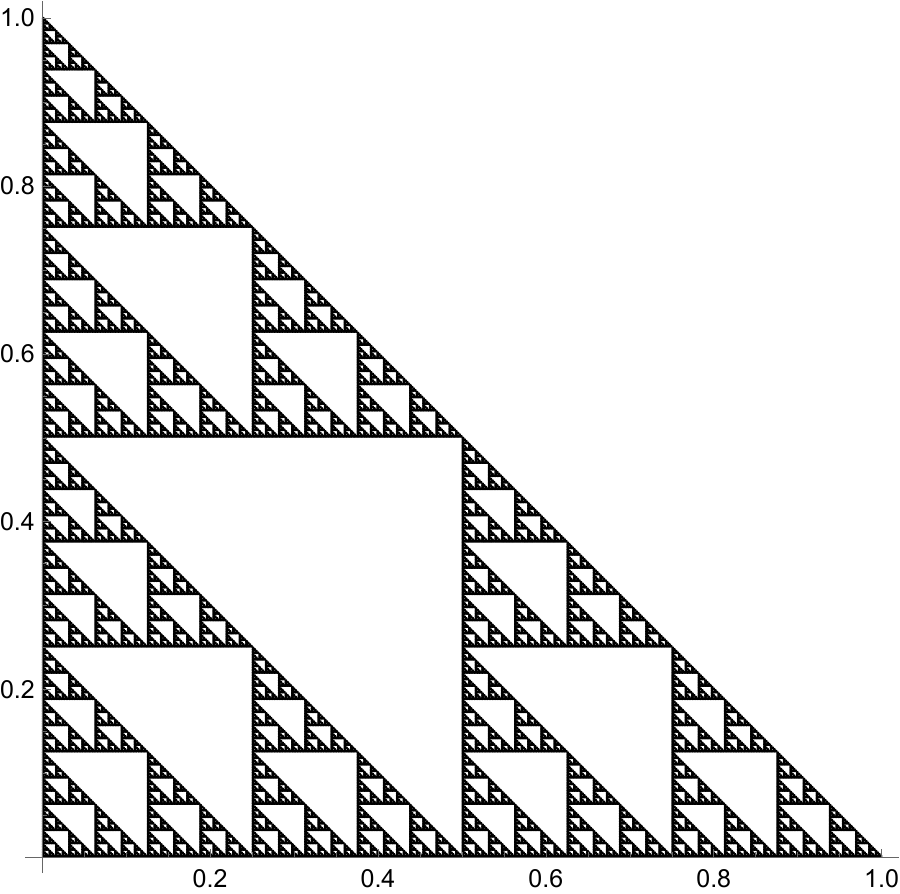}
\caption{Fractal \textit{Sierpinsky triangle} plotted by IFS algorithm.}
\label{ifs}
\end{minipage}
\end{figure}

\begin{figure}[!ht]
\begin{minipage}[b]{0.5\linewidth}
\centering
\includegraphics[scale=0.5]{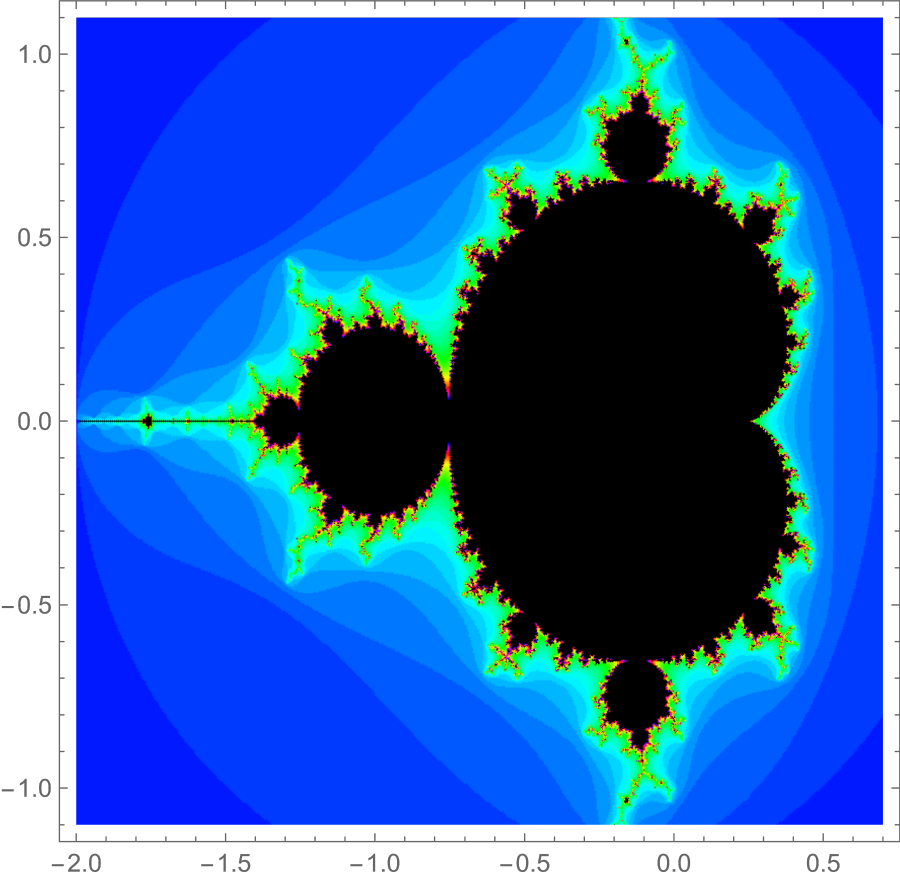}
\caption{Fractal \textit{Mandelbrott set} plotted by TEA algorithm ($z_{n + 1} = z_n^2 + c$).}
\label{teamand}
\end{minipage}
\hspace{0.5cm}
\begin{minipage}[b]{0.5\linewidth}
\centering
\includegraphics[scale=0.5]{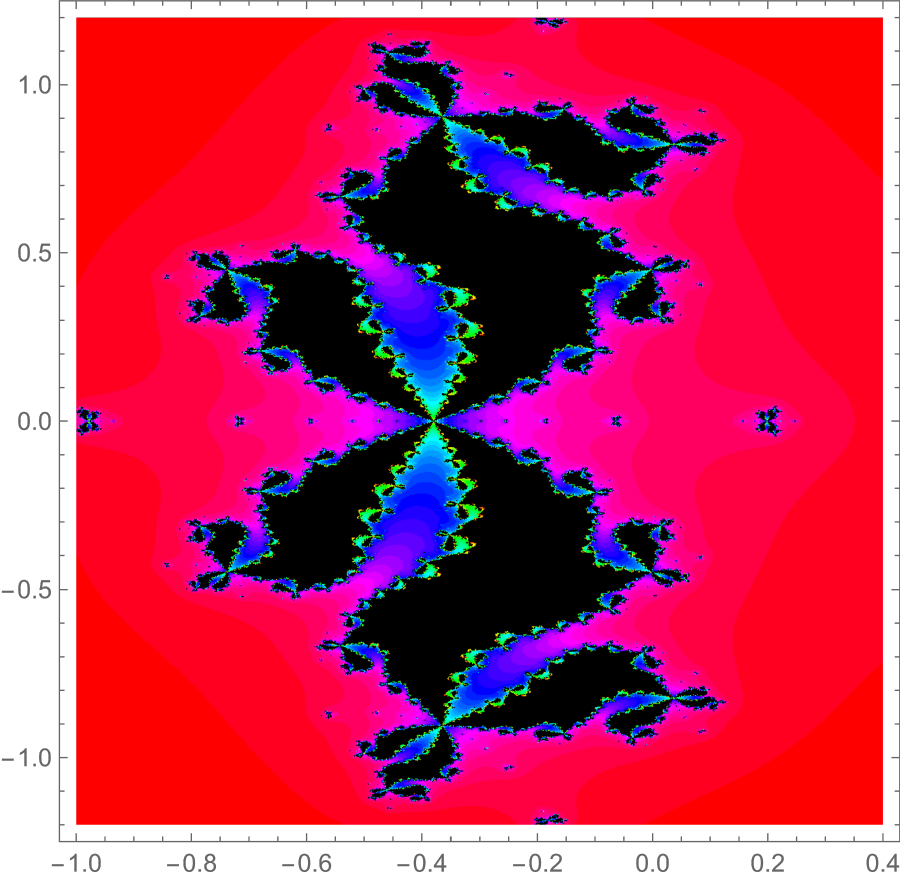}
\caption{Fractal \textit{Spider} (Julia set) plotted by TEA algorithm ($z_{n + 1} = z_n ^4 - z_n - .78)$.}
\label{tea}
\end{minipage}
\end{figure}

The main idea of our paper is to visualize computer malware. As mentioned in the previous section, visualization can be done in many ways, but they usually result in quite "boring" images with grayscale cube graphics. The question is whether visualization can be done more interestingly and simultaneously in a way that carries the necessary information. One area of mathematics has this potential and has so far been overlooked in this regard. It is the so-called fractal geometry. Fractal geometry is an area of mathematics that was officially established in the 1970s by the French mathematician B.B. Mandelbrot \cite{weibel2005mandelbrot} in the book \cite{mandelbrot1977fractal}, \cite{mandelbrot1982fractal}, \cite{mandelbrot1989fractal}. The first signs of this very special geometry appeared at the end of the 19th century when the German mathematician G. Cantor \cite{block1975noncontinuity} introduced the so-called Cantor set to the world. At this time, another German mathematician K. Weierstrass introduced to the Berlin Academy of Sciences a smooth curve with no derivative at any point \cite{weierstrass1872uber}. The other mathematical monsters (as fractals were called by mathematicians of the time) were introduced to the world, such as the so-called Sierpinski triangle \cite{whyburn1933existence}. This period was called the period of mathematical monsters and was ignored by many prominent mathematicians of the time - Ch. Hermite in a letter to T. Stieltjes: \textit{...I have turned away with horror and ugliness from that deplorable evil which is a function without derivative...}. This changed after the publication of \cite{mandelbrot1977fractal}, which clearly showed that the vast majority of dynamic events, structures and processes in our universe can straightforwardly be described just by so-called fractal geometry. Fractal geometry allows extremely complex objects to be captured very simply. In figures \ref{ifs}, \ref{tree} and \ref{teamand}, \ref{tea} are captured two examples of known fractals built using two different algorithms \cite{barnsley2014fractals}, for better imagination. There are two basic algorithms for constructing fractal objects, namely the IFS (iteration function system, generating black and white patterns) algorithm and the TEA (time escape algorithm, generating coloured patterns), \cite{barnsley2014fractals}.  

We used the TEA algorithm to visualise the malware with the construction of generative polynomials (i.e., polynomial formulas representing the image) in the manner described below. The reason for using TEA was apparent. After trivial but not obvious modification, the data obtained from the dynamic analysis of malware was suitable for constructing polynomials used by TEA to draw fractal patterns. Moreover, all fractal images are, in principle, very aesthetic in their structure and also reflect the properties of the equations that generate these images. It is a merging of pure mathematics, which in this case can visualise different types of polynomial formulas and the art arising from the nature of fractal geometry and the principle of self-repetition. We think the results of this visualisation, reported in our paper and available on GitHub\footnote{TBA if accepted}, where we offer our visualisation freely for download and further experimentation, speak for themselves.

The results of our method are \textit{malware fractals} (MF), which are presented later in this article. Fractal geometry is a fascinating and purely mathematical field about which hundreds of books have been written. We, therefore, recommend readers get a deeper understanding of fractal geometry from some of the very well-known books such as \cite{peitgen1992chaos} or \cite{barnsley2014fractals}.

\section{Malware data} 
All data for our experiments were provided by ESET\footnote{https://www.eset.com/}, which is actively collaborating with us on this research. The dataset contained a total of three file types, namely goodware, malware and potentially unwanted applications, with a total of \textbf{6,589,997} goodware, \textbf{827,853} potentially unwanted applications and \textbf{4,174,203} malware samples. 



After processing this data which included minimizing redundant entries, deleting entries that were too short, and finally shortlisting, we ended up with \textbf{4,833,893} of goodware, \textbf{421,403} of potentially unwanted applications, and \textbf{2,878,904} of malware samples for the experiment. 
For this publication, malware and goodware data were used. Potentially unwanted applications were removed from the data used. All the data obtained from ESET was generated based on dynamic analysis of malware behaviour in ESET-defined environments and on ESET servers.

\begin{figure}[!ht]
\centering
\includegraphics[scale=1]{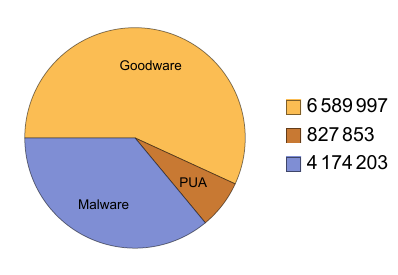}
\caption{Data-set structure from the code point of view.}
\label{p1}
\end{figure}



The malware data in this article came from a total of 164 families such as Dorkbot, Certik, Filecoder, Tinba, amongst others.


For our experiments, the most crucial part was the data capturing the use of API libraries by the malware. This part of the data package was then used to visualize the computer malware or its behaviour and subsequent image processing experiments. All the data used in this paper and our experiments were anonymized (due to NDA with ESET) for this paper by replacing the original API names with L1 and L2 up to Lx. This anonymization does not detract in any way from the veracity of the data, especially how the data captures the behaviour of the malware/goodware in question. 

The data we provide includes the generated images used for the experiments described in this paper and the iterative formulas from which those images were generated. These iterative formulas were generated based on our procedure described below.

\section{Malware fractalization} 
The core of our innovative idea is to convert the dynamic sequence of API function calls into the fractal pattern itself, done in several steps. In the first step, as mentioned earlier, the sequence of API calls of a given malware is used as recorded, with the understanding that for our experiments, these sequences were treated as sequences of vertices in the graph. The idea of displaying malware activity as a graph is not new, it has been used before, e.g. \cite{amer2020dynamic}, \cite{anderson2011graph},\cite{fan2019graph} or \cite{frenklach2021android} amongst others. The vertex sequence represents the sequential execution of API functions. To be more specific, let us say we have recorded the dataset sequences $L1, L5, L352, L4, ...$ while converting it to a graph as it has been published in our previous publications \cite{amer2020dynamic}, but also publications by other researchers, we take into account that the first activated function is L1 then L5 then L352, L4 and so on the transitions between them are characterized by arrows, $L1 \to L5 \to L352 \to L4 \to ...$, which in the graph visualization means oriented edges. This simple trick yields a graph, or if preferred, a complex network that captures the execution of API functions by the malware. This is just the first step.

\begin{figure}[!ht]
	\centering
	\includegraphics[scale=0.5]{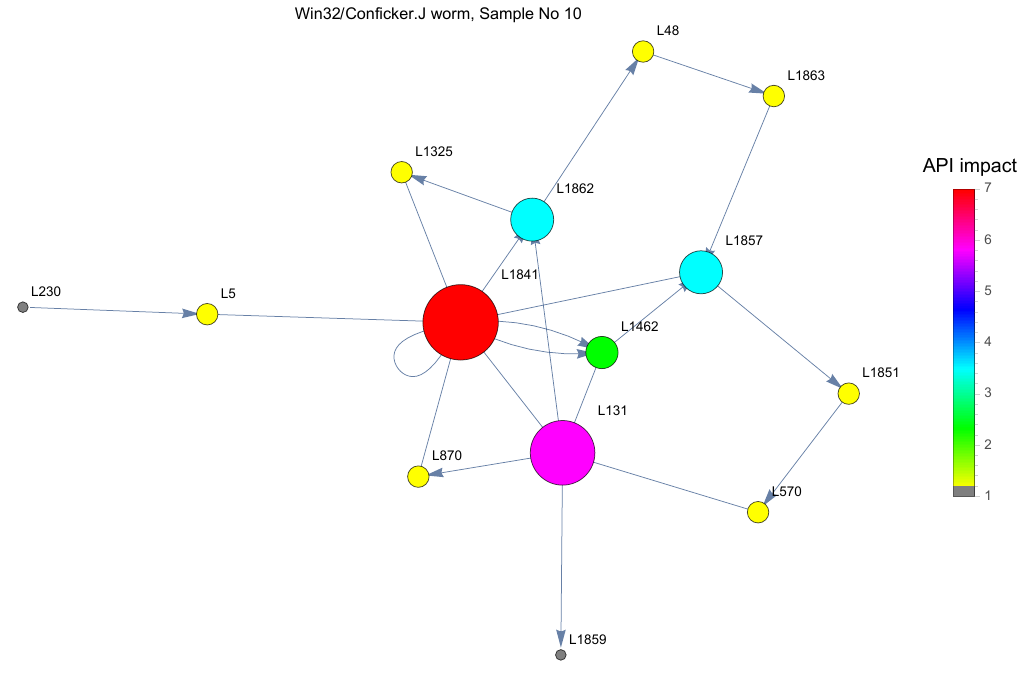}
	\caption{API call as a graph.}
	\label{API1}
\end{figure}

\begin{figure}[!ht]
	\centering
	\includegraphics[scale=0.5]{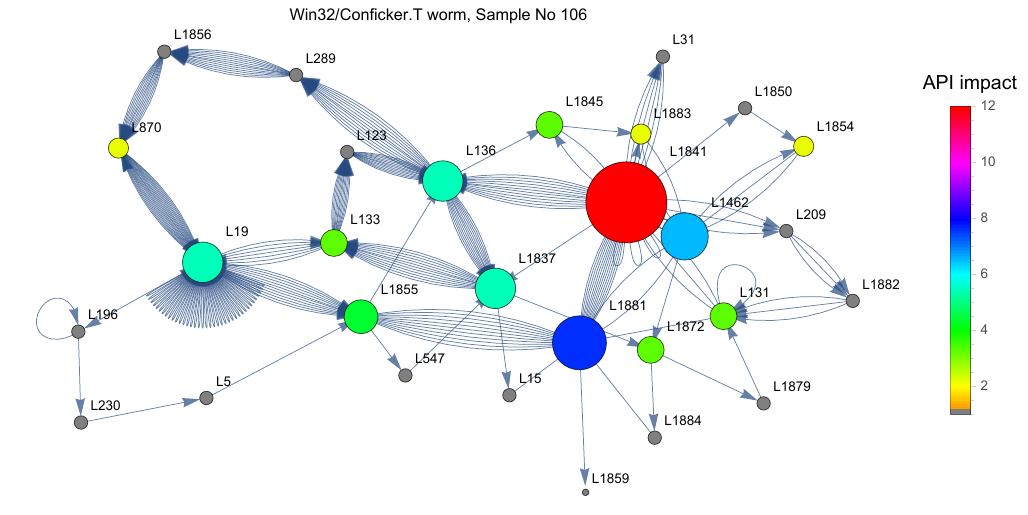}
	\caption{API call as a graph.}
	\label{API2}
\end{figure}

The resulting graphs that we obtain in this way are shown in \ref{API1} and \ref{API2}. To generate the necessary equations from these figures to generate fractal patterns that represent the behaviour of the malware, it is still essential to realize the importance of the individual vertices or API functions. Different malware uses different API functions, some repeatedly, increasing their importance within the malware functionality. This importance is shown by the colour and vertex size in the figures and captures selected centralities calculated for the graphs. In other words, we have translated the malware behaviour into a graph where the individual vertices, or API functions, have their importance. These are centralities such as \textit{degree centrality}, \textit{betweness}, \textit{eigenvector}, \textit{closeness} and others. The graph clearly shows that repeated API calls are captured here. If a call to a given API function follows several times in a row, then this is shown as an edge that comes from the same vertex it then enters (self-loop).

If we have the data in the form of a graph, the next step is step 2, where we sort the selected calculated centralities of the graph into a sequence according to Eq. (\ref{seqv}). 

\begin{equation}
	\begin{array}{l}
		\{ DC\_1,{\rm{ }}DC\_2,{\rm{ }}DC\_3,{\rm{ }}DC\_4,{\rm{ }}DC\_5,...\}  = \{ 1,{\rm{ }}5,{\rm{ }}3,{\rm{ }}2,{\rm{ }}8,...\} , \\ 
		{\rm{ }}DC{\rm{ }} = {\rm{ degree}}\,\,{\rm{centrality}} \\ 
		\{ CC\_1,{\rm{ }}CC\_2,{\rm{ }}CC\_3,{\rm{ }}CC\_4,{\rm{ }}CC\_5,...\}  = \{ {\rm{0}}{\rm{.506, 0}}{\rm{.312, 0}}{\rm{.0299, 0}}{\rm{.0981, 0}}{\rm{.0537}},...\} , \\ 
		{\rm{ }}CC{\rm{ }} = {\rm{ closeness}}\,\,{\rm{centrality}} \\ 
	\end{array}
	\label{seqv}
\end{equation}

What have we gained? We have often converted very long sequences of API calls, which can contain many API calls, into a graph with a finite number of vertices in the order of units to tens. Here, the repeatability of the call in the original record is only reflected by increasing the number of edges or the weight of the edge between two vertices but also by the importance of the vertex (i.e., its size and colour). This brings with it the advantage that one can compute the significance of such API functions for a given graph or, if you like, for the dynamics of malware.

Obtaining a mathematical prescription for generating MF is, in principle, already very simple. However, the procedure presented here is not the only one possible. The procedures and results presented here result from more than a year of research, where we have tried both logical analysis and trial-and-error to find the best way to rewrite the API sequence map into a mathematical relation that allows us to generate a given fractal pattern. This, then, more or less represents the structure of the malware behaviour. The sequence from \ref{seqv} is then used to multiply and amplify the variable $z$ (understood as a complex variable in fractal geometry). The principle is illustrated in Eq. \ref{mf}.

\begin{equation}
	\begin{array}{l}
		generally \\
		z_{n + 1}=CC\_1 z_{n}^{DC\_1} + CC\_2 z_{n}^{DC\_2} + CC\_3 z_{n}^{DC\_3} + CC\_4 z_{n}^{DC\_4} + CC\_5 z_{n}^{DC\_5} ... \\
		\\
		{\rm{exactly}} \\
		z_{n + 1}=0.506z_{n} + 0.312z_{n}^5 + 0.0299z_{n}^3 + 0.0981z_{n}^2 + 0.0537z_{n}^8 ... \\
	\end{array}
	\label{mf}
\end{equation}

It is, therefore, a kind of modification of the Mandelbrot set. If we add some centralities as a constant shift in the sense of the $x$ and $y$-axis, which could also come from the computed centralities, then we get into the region of the Julia sets \cite{barnsley2014fractals}, Fig. \ref{tea}, of which there is also a plethora. This modification is one possible topic for future research in this direction.

\begin{figure}[!ht]
\centering
\includegraphics[scale=0.35]{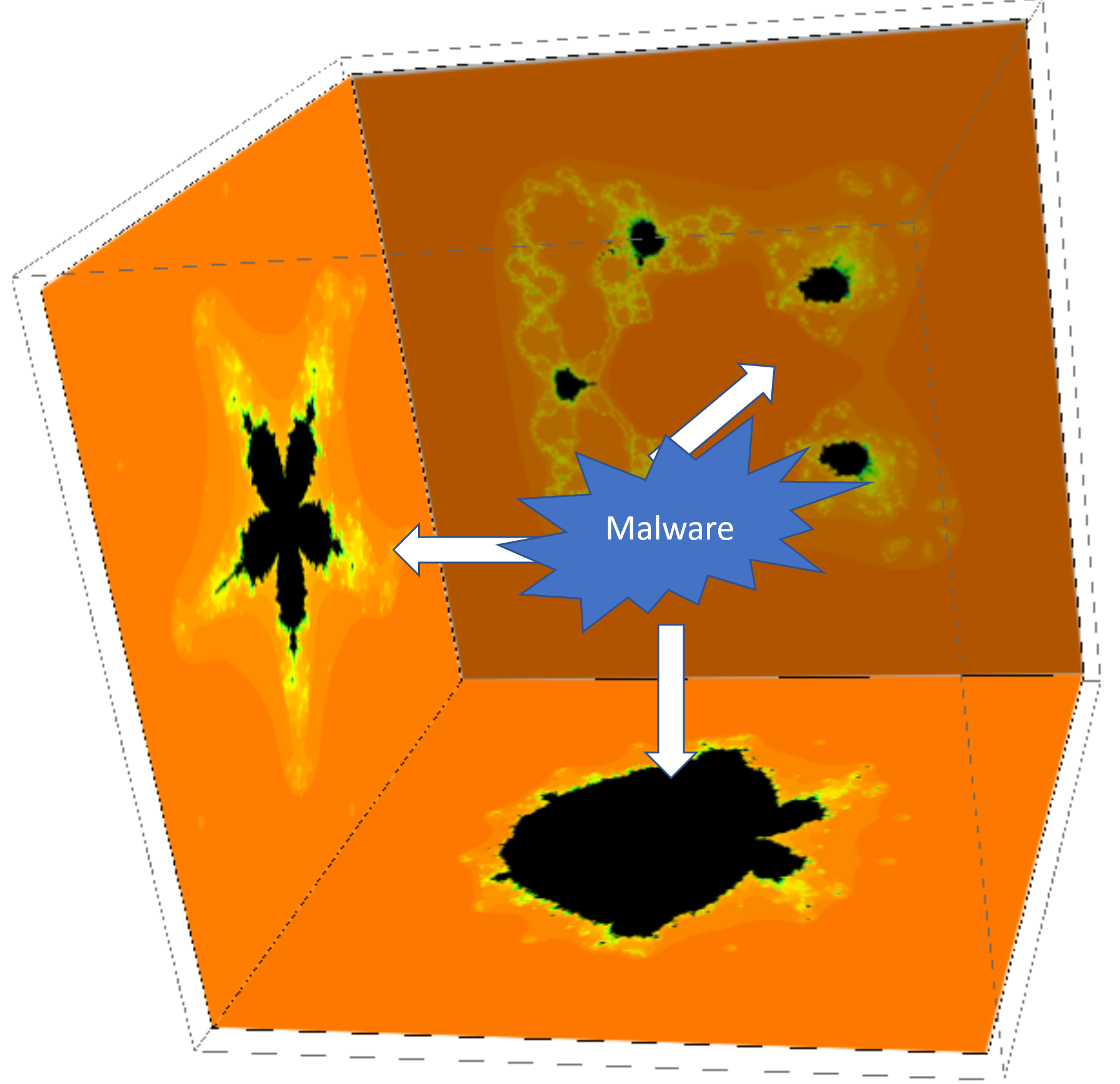}
\caption{Malware as the multidimensional fractal? 3 faces of one malware object.}
\label{3D}
\end{figure}

\begin{figure}[!ht]
\begin{minipage}[b]{0.5\linewidth}
\centering
\includegraphics[scale=0.3]{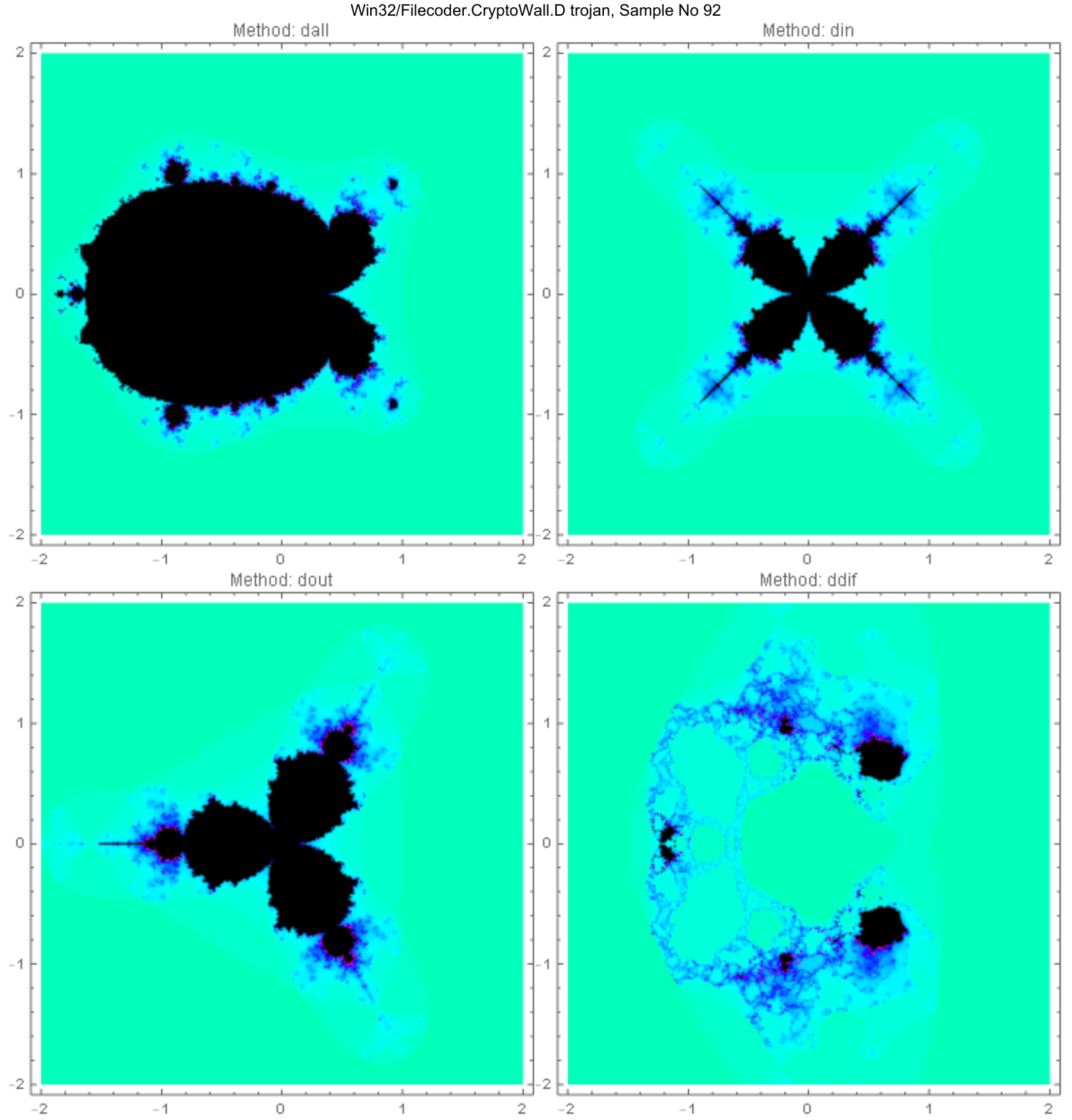}
\caption{Malware Cryptowall as a fractal object.}
\label{Cryptowall}
\end{minipage}
\hspace{0.5cm}
\begin{minipage}[b]{0.5\linewidth}
\centering
\includegraphics[scale=0.3]{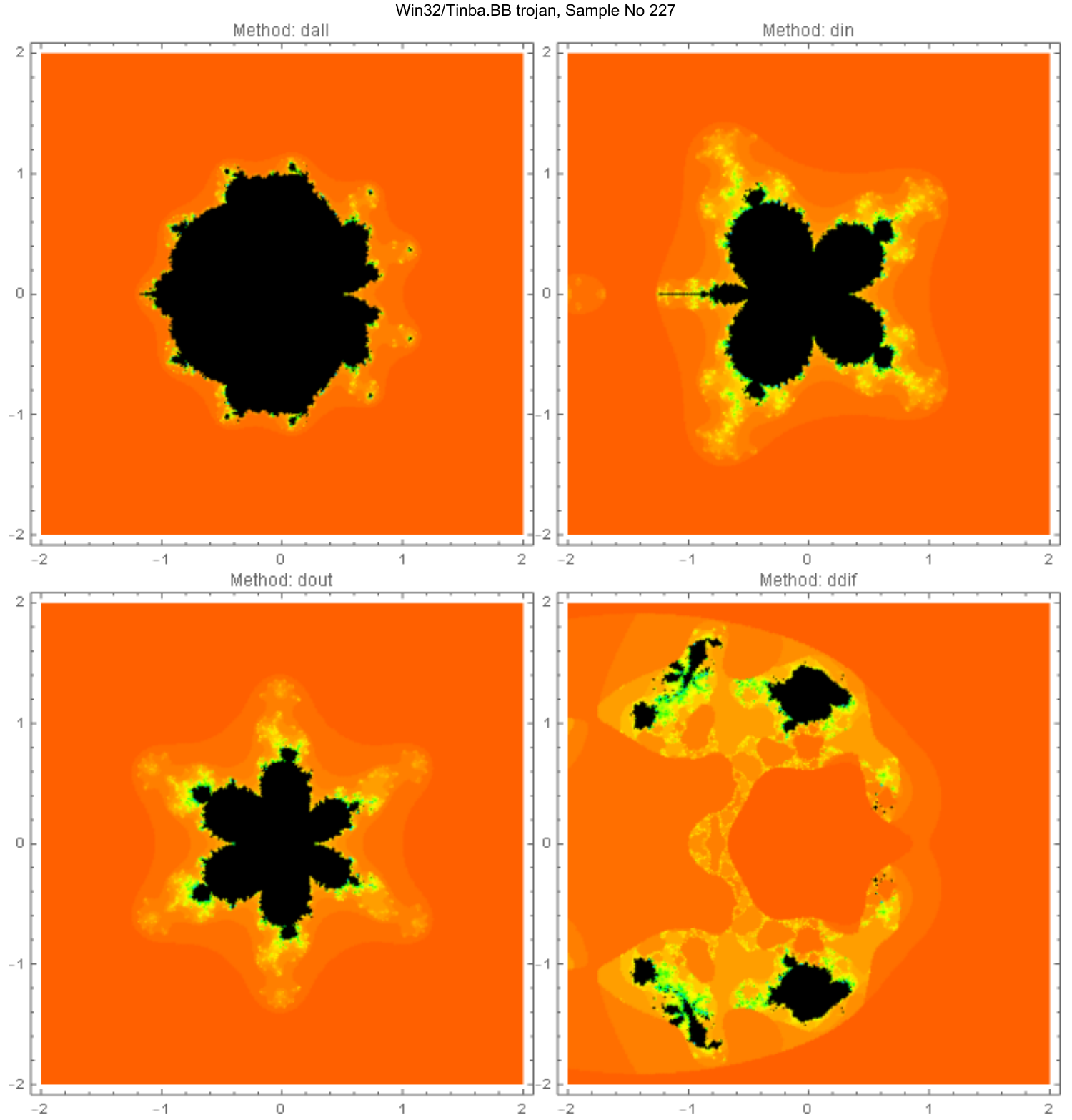}
\caption{Malware Tinba as a fractal object.\\}
\label{Tinba}
\end{minipage}
\end{figure}

\begin{figure}[!ht]
\begin{minipage}[b]{0.5\linewidth}
\centering
\includegraphics[scale=0.3]{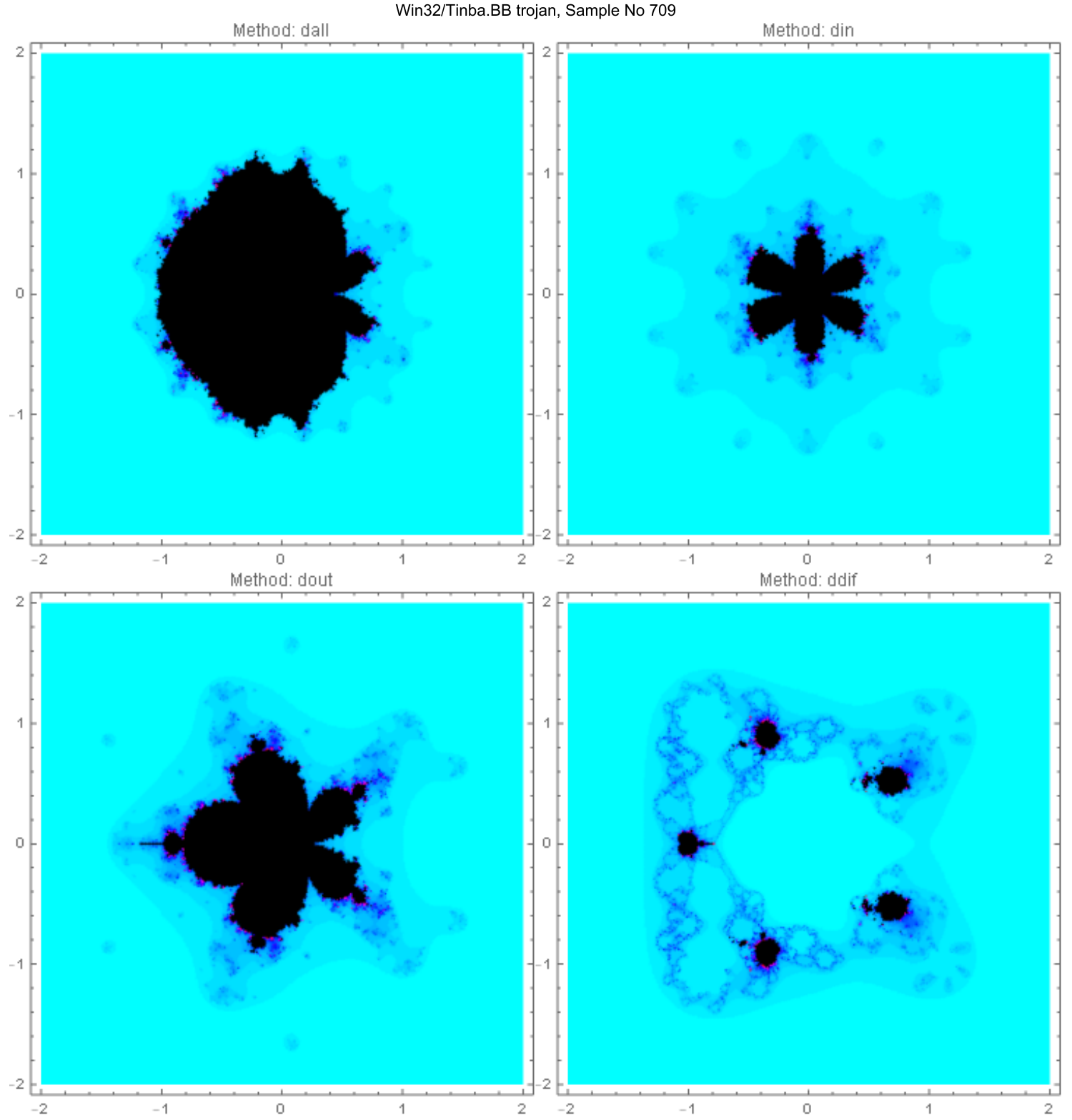}
\caption{Malware Tinba in the form of a fractal object.}
\label{Tinba2}
\end{minipage}
\hspace{0.5cm}
\begin{minipage}[b]{0.5\linewidth}
\centering
\includegraphics[scale=0.3]{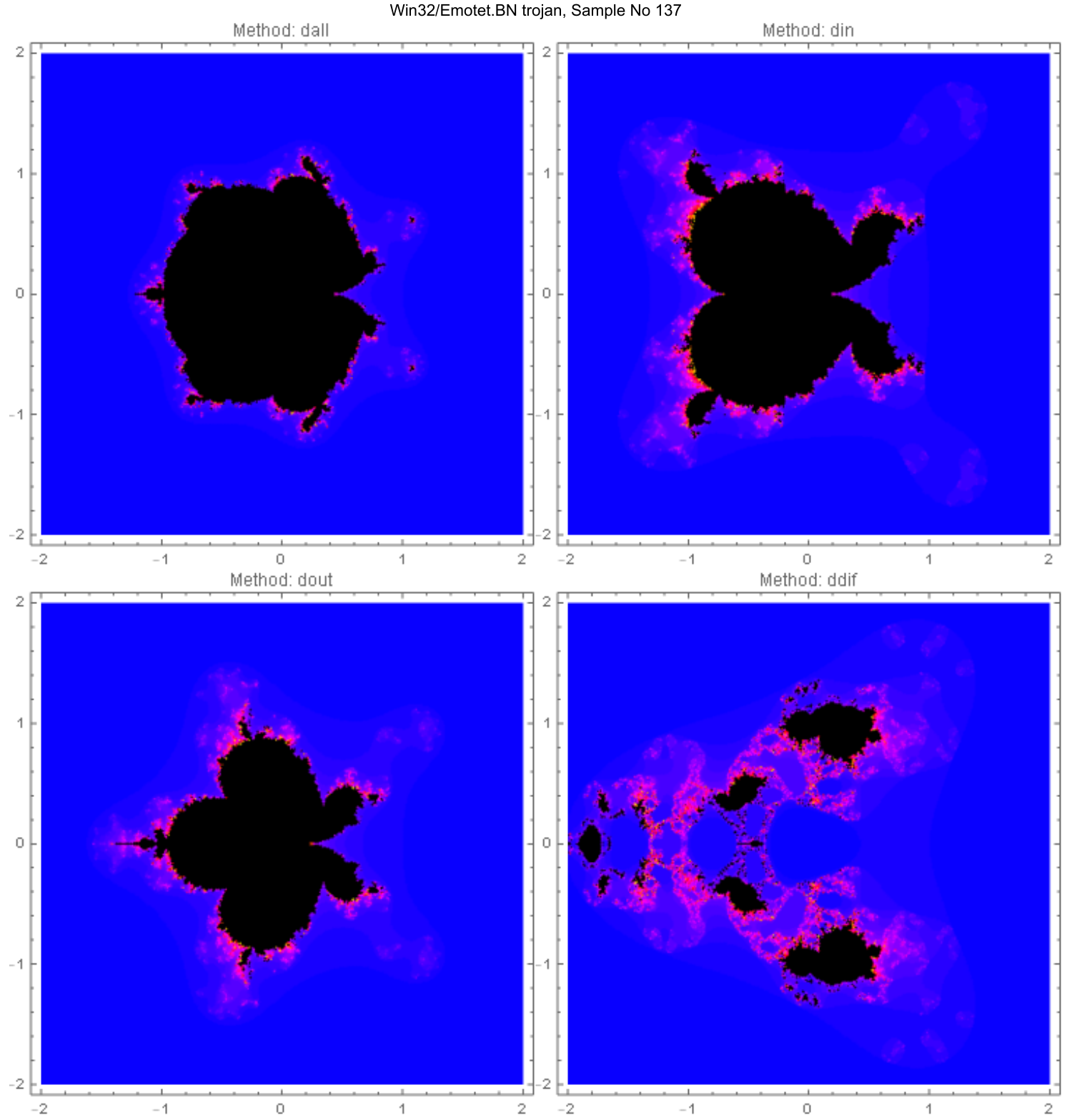}
\caption{Malware Emotet as a fractal object.}
\label{Emotet}
\end{minipage}
\end{figure}

\begin{figure}[!ht]
\begin{minipage}[b]{0.5\linewidth}
\centering
\includegraphics[scale=0.3]{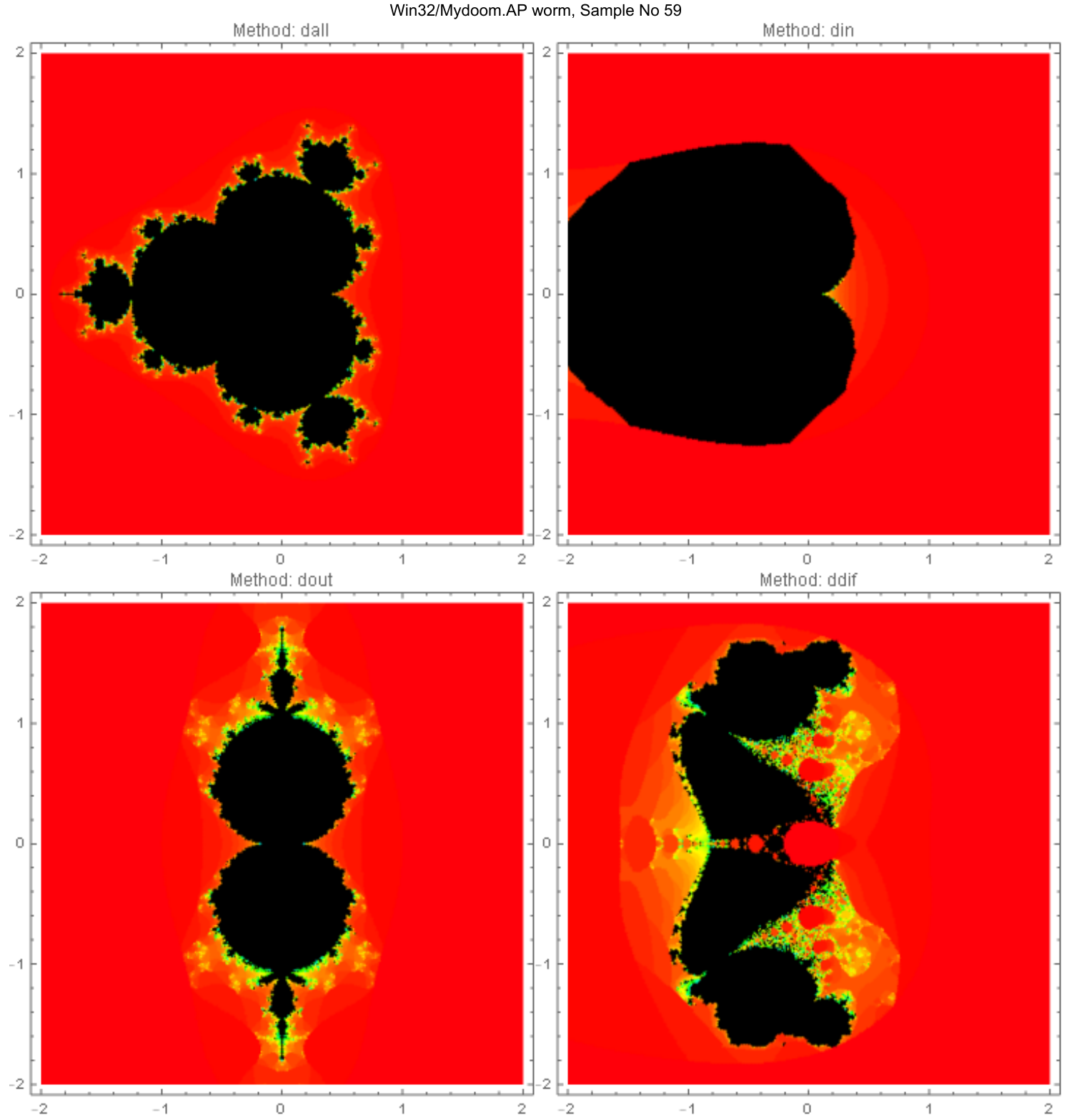}
\caption{Malware Mydoom as a fractal object.}
\label{Mydoom}
\end{minipage}
\hspace{0.5cm}
\begin{minipage}[b]{0.5\linewidth}
\centering
\includegraphics[scale=0.3]{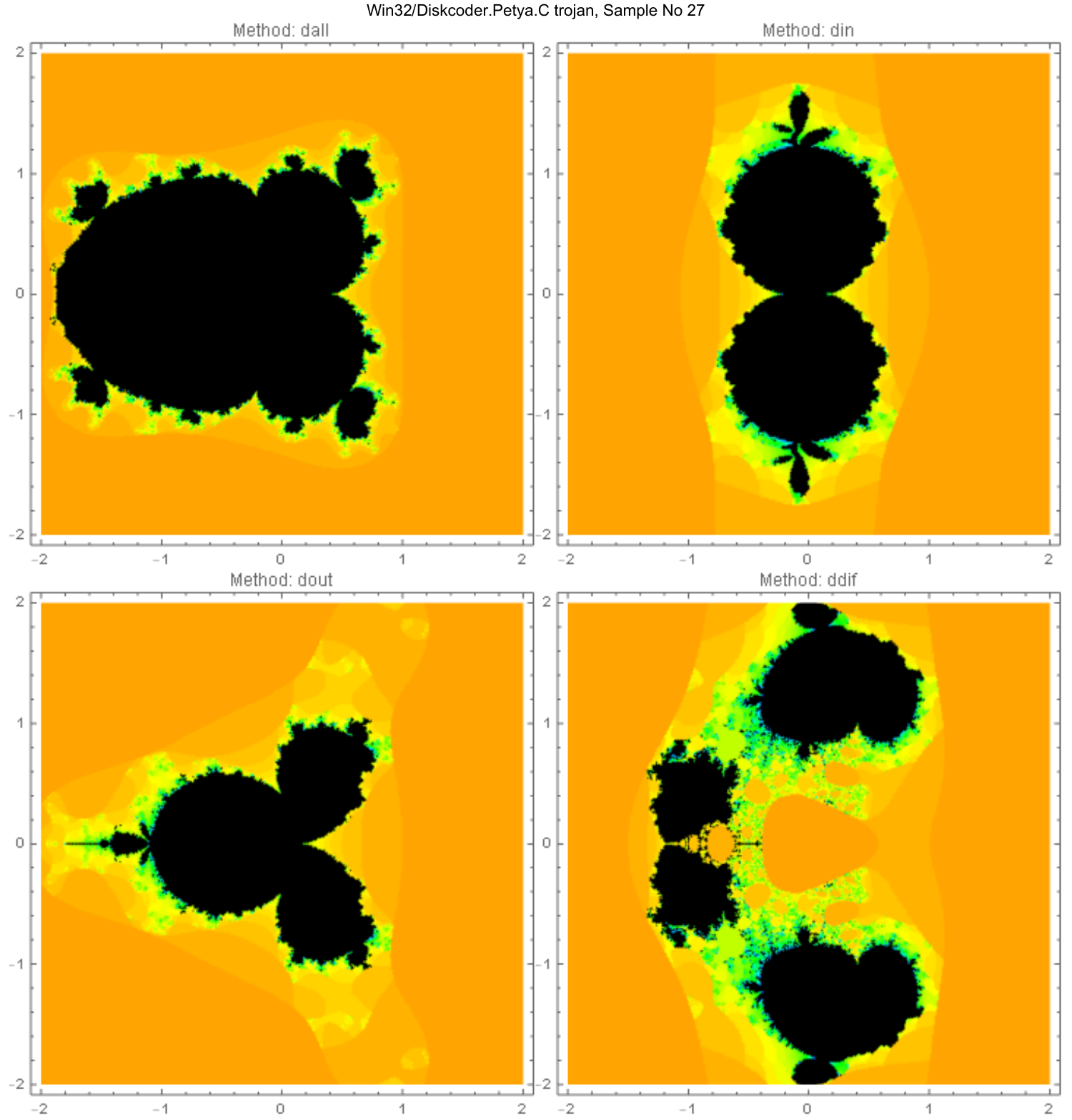}
\caption{Malware Petya as a fractal object.\\}
\label{Petya}
\end{minipage}
\end{figure}

The last step is to obtain an image of the malware fractal. The actual generation of the malware fractal is straightforward after creating the iterative formulas. According to the \cite{barnsley2014fractals} algorithms and the policy for generating Mandelbrot sets or Julia sets, malware fractals are generated that reflect the dynamics of the behaviour of the malware in their structure and colour. The actual generation is iterative - the TEA algorithm \cite{barnsley2014fractals}, which uses the appropriate equation corresponding to the corresponding visualization of the malware behaviour. Figures \ref{Cryptowall} - \ref{Mydoom} show examples of some of the visualizations that have been generated. 
All the images, including the supporting data (generating equations and FractalVizualizer software), are available on the GitHub\footnote{TBA} repository. During our experimentation, we found that of the possible interesting descriptions based on different centralities, each showing the malware slightly differently, our approach is the most promising. Therefore, we decided to generate quartets of images (degree centralities - In + Out, In, Out, In - Out) that present the malware in question in this way, figuratively speaking, from different perspectives. These quartets figures \ref{Cryptowall} - \ref{Mydoom} can be analogously understood as projections of possible malware object representations from $N$-dimensional space onto a 2D plane. Thus we get a few different views on malware behaviour - i.e. its "dynamical behaviour fingerprint". This idea is demonstrated in Figure \ref{3D}.
The principle of generating a fractal representation of malware is also sketched in the pseudocode Algorithm \ref{MFAlg}.

\clearpage
\RestyleAlgo{ruled}

\begin{algorithm}[!ht]
	\SetKwComment{Comment}{/* }{ */}
	\SetKwInOut{Input}{input}\SetKwInOut{Output}{output}
	\caption{Malware fractalization pseudocode}
	\label{MFAlg}
	\Input{API sequence call, parameters for TEA (e.g. image range, number of iterations to calculate escaping trajectory)}
	\Output{4 quadrant figure of malware fractal, as for example on \\Fig. \ref{Cryptowall}}
	\Comment{Preprocess API sequences}
	[API selection] $\to$ API sequence collecting from database records\;
	\Comment{Polynomial iterative formulas for fractal images creation, $z$ is the complex value}
	[Graph synthesis] $\to$ API sequence is converted into a graph, e.g. Fig. \ref{API1} or \ref{API2}\;
	[Graph centrality calculation] $\to$ For all graph vertice's degree centrality are calculated and four sequences for four quadrant are generated (see Eq. \ref{mf}): 
	\begin{enumerate}
		\item \textbf{InDegree} (for incomming connections, DIn), Q1, e.g. $z_{n+1}=0.631 z_{n}^2 + 0.099 z_{n}^3 + 0.268 z_{n}^4$
		\item \textbf{Degree} (Dall), Q2, e.g. $z_{n+1}=0.631 z_{n} + 0.368 z_{n}^2$
		\item \textbf{OutDegree} (for outcomming, DOut), Q3, e.g. $z_{n+1}=0.731 z_{n} + 0.268 z_{n}^5$
		\item \textbf{Ddiff} = DIn-DOut, Q4, e.g. $z_{n+1}=0.900 z_{n}^2 + 0.099 z_{n}^6$
	\end{enumerate}
	\Comment{Figures generation, using TEA, see \cite{barnsley2014fractals}}
	\While{ all 4 quadrant images are not calculate}{
		\begin{enumerate}
			\item {Escaping trajectory is calculated: take the first formula and for each (starting) $z_{0}$ = point in plane (e.g. $Re(z)$ and $Im(z)$ $\in$ $[-2, 2]$, is iterative trajectory calculated).}
			\item {Check perimeter for escaping (this case set to 2) by $Abs[z_{n}]$.}
			\item {Colorize. If $Abs[z_{n}] < 2$ then continue in iterations else set starting point by color related to number of used iterations and move to the next plane point (the trajectory has escaped after $n$ iterations); if trajectory in all iterations do not escape, set starting point Black.}
			\item {When all plane points are calculated for escaping trajectory, take the next formula and calculate new quadrant figure.}
		\end{enumerate}
	}
\end{algorithm}

The four equations used in this algorithm were chosen based on further experimentation with different centralities and their possible uses to construct polynomial formulas generating fractal patterns. The idea was, for example, that the $z$ powers have to be integers, which a-priori is satisfied by degree centrality. For example, closeness centrality could be used as a multiplication factor for a given $z$. Of course, other possible combinations are not excluded; this topic is open for further research.

Our method raises the question of what can be considered a bottleneck and whether there is a partial loss of information. The bottleneck is the conversion of API sequences to a given graph because there is a non-zero probability that other malware will generate a graph with the same structure (e.g. different API calls but the same graph structure). But this is a problem with all malware classification approaches that use visualization in the form of an image. The conversion to a fractal image itself does not lose information. The graph, or the resulting polynomial formula, can be considered a compression representation of the fractal \cite{barnsley2014fractals}. From this point of view, see Banach's theorem \cite{barnsley2014fractals}, the fractal representation of malware can be seen as a decompression of the information contained in the graph representation. The question then is how much information is obtained by the fractal representation and how it can be interpreted. 

Our initial goal was to convert the API calls to fractals as a new unique visualization attempt. Then we apply classification. This conversion to fractals is functional, or rather the fractal images still carry enough information - the convolutional neural network can train over these images and achieves well-acceptable accuracy. As a result, we know that the fractal images already carry enough information at this stage because the convolutional neural network can categorize.

\section{Experiment structure} 
\begin{figure}[!ht]
\begin{minipage}[b]{0.5\linewidth}
\centering
\includegraphics[scale=0.13]{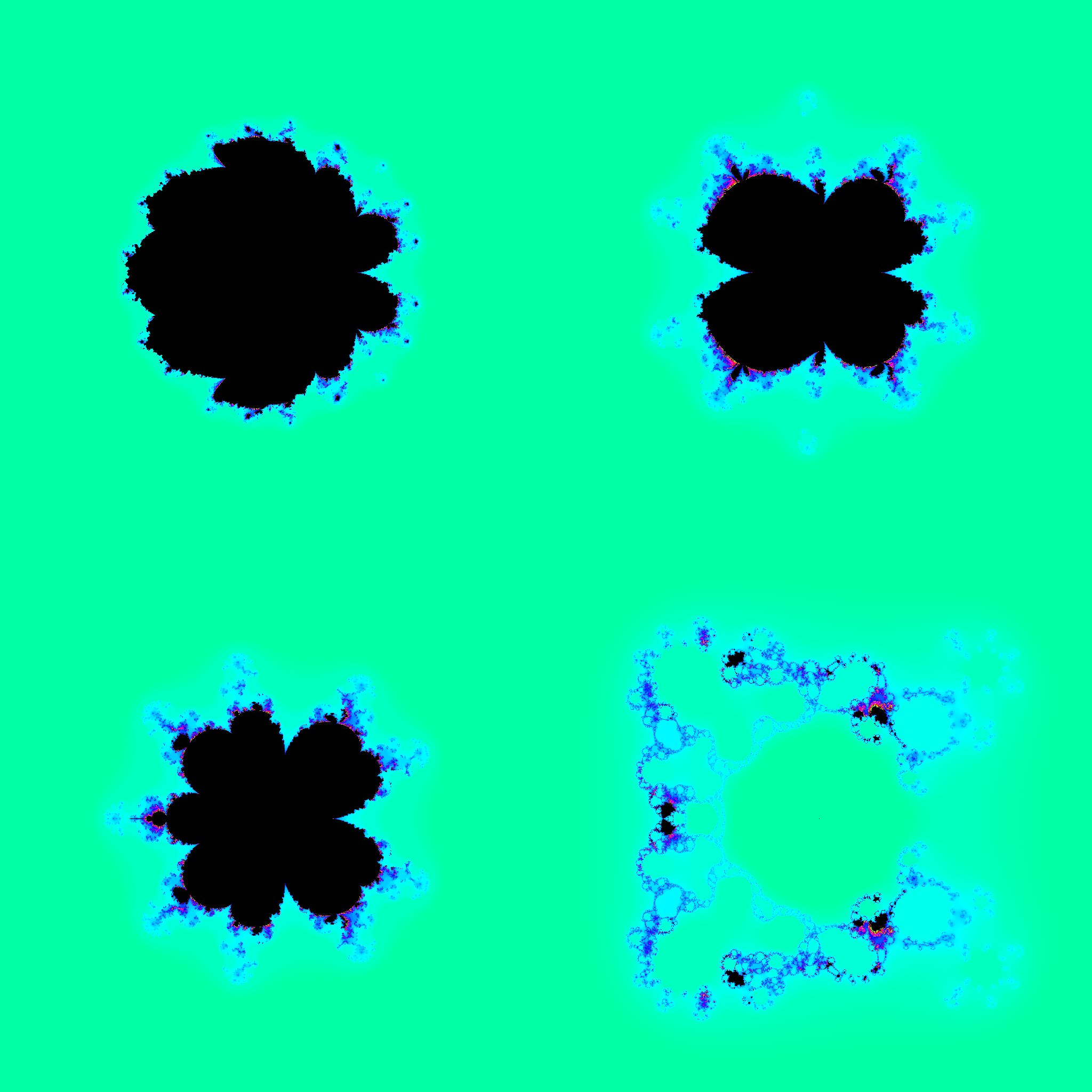}
\caption{Malware Win32.Adware.BrowseFox as a fractal object; fixed number of iterations used.}
\label{Win32.Adware.BrowseFox}
\end{minipage}
\hspace{0.5cm}
\begin{minipage}[b]{0.5\linewidth}
\centering
\includegraphics[scale=0.13]{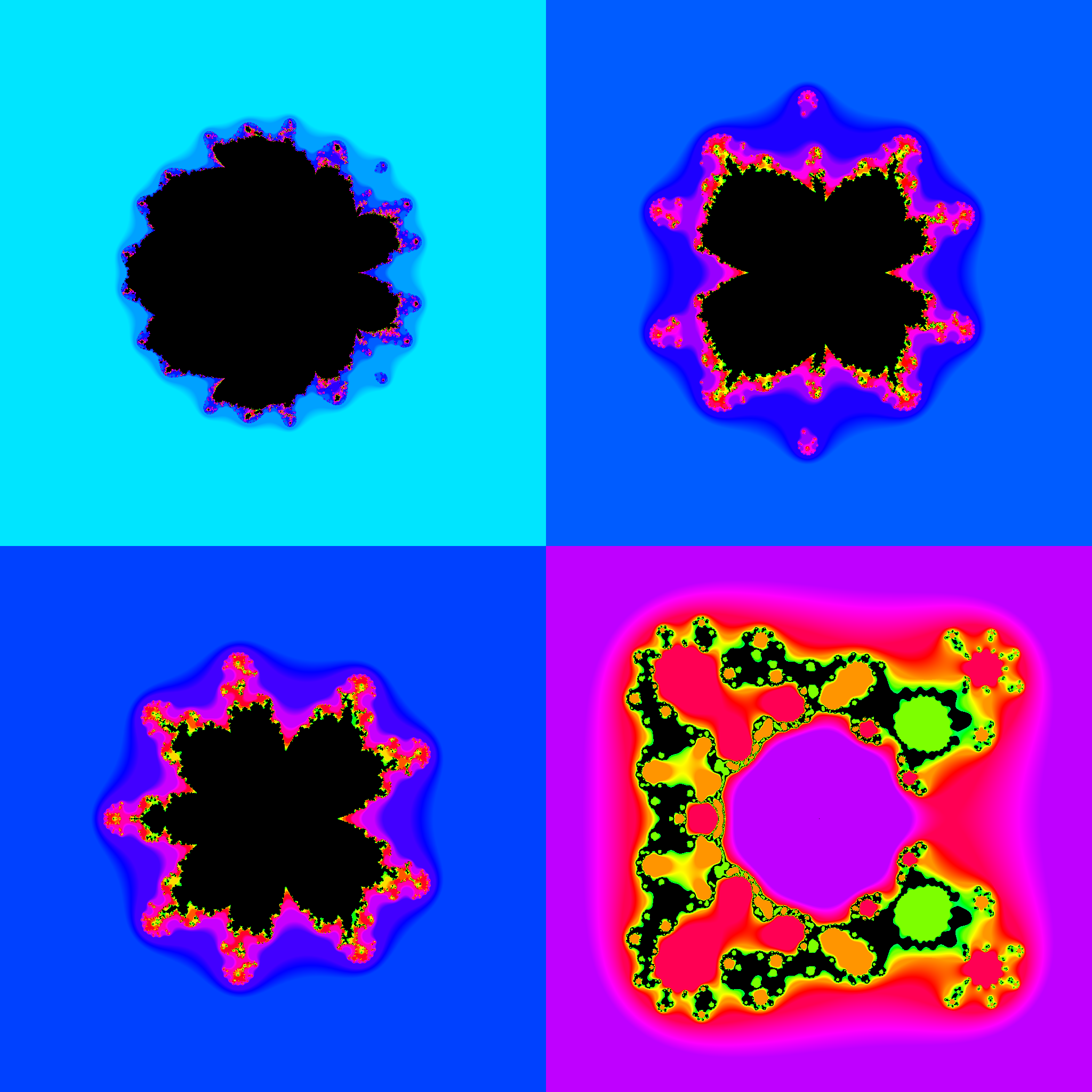}
\caption{Malware Win32.Adware.BrowseFox as a fractal object; customized number of iterations used.}
\label{Win32.Adware.BrowseFoxExt}
\end{minipage}
\end{figure}

To verify our idea, we performed experiments 1) analysis of the dependence of classification quality on image resolution (see section \ref{dp}) and 2) malware vs goodware with a modified deep learning network. This was done on randomly selected data from the ESET database with total robustness of \textbf{4,833,893} goodware samples and \textbf{2,878,904} malware samples.
Both experiments used a total of 65,270 images to distinguish between goodware (33,173) and malware (32,097) pieced together randomly from all 164 malware families. The experiments were repeated over two image sets (see figures \ref{Win32.Adware.BrowseFoxExt} and \ref{Win32.Adware.BrowseFox}, see section \ref{MGF}) with different graphical concepts. Thus a total of over 130,540 images were used. A deep-learning network processed these image sets. The results of each experiment are described below. Python was used as the software with the corresponding Tensorflow 2.0 libraries. All this was run on a standard desktop computer.

\section{Data preprocessing} 
\label{dp}
In this section, we briefly review how we investigated the effect of image resolution on the classification quality and possible modification of the malware representation into a fractal pattern, where we tried to bring more information into its structure to make the MF more distinguishable.

\subsection{Image resolution}
All data for the experiment were cleaned of records containing minor uncertainties. Similarly, duplicate records were deleted to prepare the data for generating unique and unambiguous images. By deleting duplicate records, we mean records that were the same. However, the same malware types with different analysis record lengths are present in the database. So naturally, we left such records in the database and used them to render the images. Next, we mention the preparation of the image database and the preliminary findings we found when checking the fractal images of the malware (Sec. \ref{mutmal}).

Before the experiment, we addressed how much image resolution is optimal for classification. To this end, we would run a number of experiments for different resolution levels, and the results are shown in Table \ref{tbl}. It can be seen that higher resolution makes sense, but it is very computationally intensive. 

For example:
\begin{itemize}
    \item Resolution 768$x$768, batch size=16 $\to$ 39 GB RAM 
    \item Resolution 512$x$512, batch size=16 $\to$ 20 GB RAM 
\end{itemize}

\begin{table}[!ht]
\caption{Classification dependence on image resolution. NTP - number of trained parameters. By cyan and green color are marked the most interesting results.}
\label{tbl}
\centering
\begin{tabular}{lcc|ccc|r}
\rowcolor[HTML]{EFEFEF} 
\hline
Resolution                              & \multicolumn{1}{l}{\cellcolor[HTML]{EFEFEF}Epoch} & \multicolumn{1}{l}{\cellcolor[HTML]{EFEFEF}Batch Size} & \multicolumn{1}{l}{\cellcolor[HTML]{EFEFEF}Training} & \multicolumn{1}{l}{\cellcolor[HTML]{EFEFEF}Validation}        & \multicolumn{1}{l}{\cellcolor[HTML]{EFEFEF}Test}              & \multicolumn{1}{l}{\cellcolor[HTML]{EFEFEF}NTP} \\
\hline
\rowcolor[HTML]{CFE2F3} 
\textbf{1024x1024}                      & 5                                                 & 8                                                      & 98                                                   & \cellcolor[HTML]{00FFFF}\textbf{84.54}                        & \cellcolor[HTML]{00FF00}\textbf{82.69}                        & 134 313 186                                     \\
\rowcolor[HTML]{CFE2F3} 
\textbf{1024x1024}                      & 10                                                & 8                                                      & 98.63                                                & \cellcolor[HTML]{00FFFF}\textbf{82.47}                        & \cellcolor[HTML]{00FF00}\textbf{82.69}                        & 134 313 186                                     \\
\hline
\hline
\rowcolor[HTML]{D9EAD3} 
\textbf{768x768}                        & 5                                                 & 16                                                     & 97.54                                                & \cellcolor[HTML]{00FFFF}\textbf{83.51}                        & 80.7                                                          & 75 592 930                                      \\
\rowcolor[HTML]{D9EAD3} 
\textbf{768x768}                        & 10                                                & 16                                                     & 99.66                                                & \cellcolor[HTML]{00FFFF}\textbf{81.96}                        & \cellcolor[HTML]{00FF00}\textbf{84.61}                        & 75 592 930                                      \\
\rowcolor[HTML]{D9EAD3} 
{\color[HTML]{0000FF} \textbf{512x512}} & {\color[HTML]{0000FF} \textbf{5}}                 & {\color[HTML]{0000FF} \textbf{16}}                     & {\color[HTML]{0000FF} \textbf{95.53}}                & \cellcolor[HTML]{00FFFF}{\color[HTML]{0000FF} \textbf{81.44}} & \cellcolor[HTML]{00FF00}{\color[HTML]{0000FF} \textbf{84.61}} & {\color[HTML]{0000FF} \textbf{33 649 890}}      \\
\rowcolor[HTML]{D9EAD3} 
\textbf{512x512}                        & 10                                                & 16                                                     & 99.03                                                & \cellcolor[HTML]{00FFFF}\textbf{81.44}                        & 75                                                            & 33 649 890                                      \\
\rowcolor[HTML]{D9EAD3} 
\textbf{512x512}                        & 5                                                 & 32                                                     & 94.62                                                & 80.41                                                         & \cellcolor[HTML]{00FF00}\textbf{84.61}                        & 33 649 890                                      \\
\rowcolor[HTML]{D9EAD3} 
\textbf{512x512}                        & 10                                                & 32                                                     & 99.31                                                & \cellcolor[HTML]{00FFFF}\textbf{82.47}                        & 82.69                                                         & 33 649 890                                      \\
\rowcolor[HTML]{D9EAD3} 
\textbf{384x384}                        & 5                                                 & 16                                                     & 95.36                                                & 78.35                                                         & 80.7                                                          & 18 969 826                                      \\
\rowcolor[HTML]{D9EAD3} 
\textbf{384x384}                        & 10                                                & 16                                                     & 99.77                                                & 80.93                                                         & 80.7                                                          & 18 969 826                                      \\
\rowcolor[HTML]{D9EAD3} 
\textbf{256x256}                        & 5                                                 & 16                                                     & 91.81                                                & 79.9                                                          & 76.9                                                          & 8 484 066                                       \\
\rowcolor[HTML]{D9EAD3} 
\textbf{256x256}                        & 10                                                & 16                                                     & 98.57                                                & 80.93                                                         & 80.7                                                          & 8 484 066                                       \\
\hline
\hline
\rowcolor[HTML]{FCE5CD} 
\textbf{192x192}                        & 5                                                 & 16                                                     & 89.92                                                & 78.87                                                         & 76.9                                                          & 4 814 050                                       \\
\rowcolor[HTML]{FCE5CD} 
\textbf{192x192}                        & 10                                                & 16                                                     & 98.4                                                 & 79.38                                                         & 75                                                            & 4 814 050                                       \\
\rowcolor[HTML]{FCE5CD} 
\textbf{128x128}                        & 5                                                 & 16                                                     & 88.09                                                & 79.38                                                         & 73.08                                                         & 2 192 610                                       \\
\rowcolor[HTML]{FCE5CD} 
\textbf{128x128}                        & 10                                                & 16                                                     & 97.82                                                & 80.93                                                         & 73.08                                                         & 2 192 610                                       \\
\rowcolor[HTML]{FCE5CD} 
\textbf{64x64}                          & 5                                                 & 16                                                     & 84.31                                                & 76.29                                                         & 73.08                                                         & 619 746                                         \\
\rowcolor[HTML]{FCE5CD} 
\textbf{64x64}                          & 10                                                & 16                                                     & 95.76                                                & 77.84                                                         & 75                                                            & 619 746                                         \\
\rowcolor[HTML]{FCE5CD} 
\textbf{32x32}                          & 5                                                 & 16                                                     & 82.93                                                & 76.29                                                         & 78.84                                                         & 226 530                                         \\
\rowcolor[HTML]{FCE5CD} 
\textbf{32x32}                          & 10                                                & 16                                                     & 96.22                                                & 81.44                                                         & 73                                                            & 226 530
\end{tabular}
\end{table}

As far as the accuracy (Tab. \ref{tbl}) itself is concerned, it is pretty clear that it is definitely acceptable, and the method we have presented hides a great potential for further improvement. One can also take into account the claims\footnote{https://www.obviously.ai/post/machine-learning-model-performance}

\textit{Good accuracy in machine learning is subjective. But in our opinion, anything greater than 70\% is a great model performance. In fact, an accurate measure of anything between 70\%-90\% is not only ideal, but it is also realistic. This is also consistent with industry standards.}

next\footnote{https://stephenallwright.com/good-accuracy-score/}

\textit{Over 90\% - Very good, between 70\% and 90\% - Good, between 60\% and 70\% - OK, below 60\% - Poor}

or opinion\footnote{https://towardsdatascience.com/should-you-continue-improving-the-accuracy-of-your-machine-learning-model-5ad9c63c796b}

\textit{It is possible to build a good ML algorithm with 80\%–85\% of accuracy using suitable techniques; however, to achieve a better accuracy (85\%–95\%), it takes a significant amount of time, effort, deeper domain knowledge, extreme data engineering, more data collection, and so on.}

From Table \ref{tbl}, it can be seen that the optimal resolution is 512$\times$512 pixels (768$\times$768 has similar results, but due to the computational and memory requirements, it seemed optimal to use 512$\times$512.), but it should be noted that this statement only applies to the deep learning network configuration we used given in Tab. \ref{tabdl}, 
where We created a simple convolutional neural network model for the experiments. The network consists of convolutional layers, with pooling occurring after each layer. After the last convolutional layer and pooling, there is a dropout layer to help against overtraining the model. After the dropout layer, a flattened layer converts the data into 1D to make this input suitable for an already classical, fully connected neural network. Precisely, the flattened layer is followed by two fully connected layers and an output layer. More detailed information can be seen in the model summary in Tab. \ref{tabdl}.
It is clear that higher resolution carries more information; however, it is very memory intensive, as well as the overall network configuration designed for deep learning.

\begin{table}[!ht]
\centering
\caption{Used model.}
\small
\begin{tabular}{lll}
\hline
Layer (type)                      & Output Shape          & Param \#   \\
\hline
rescaling\_2 (Rescaling)          & (None, 512, 512, 3)   & 0          \\
max\_pooling2d\_3 (MaxPooling 2D  & (None, 256, 256, 32)  & 0          \\
conv2d\_4 (Conv2D)                & (None, 256, 256, 64)  & 18496      \\
max\_pooling2d\_4 (MaxPooling2D)  & (None, 128, 128, 64)  & 0          \\
conv2d\_5 (Conv2D)                & (None, 128, 128, 128) & 73856      \\
max\_pooling2d\_5 (MaxPooling 2D) & (None, 64, 64, 128)   & 0          \\
dropout\_1 (Dropout)              & (None, 64, 64, 128)   & 0          \\
flatten\_1 (Flatten)              & (None, 524288)        & 0          \\
dense\_3 (Dense)                  & (None, 64)            & 33554496   \\
dense\_4 (Dense)                  & (None, 32)            & 2080       \\
dense\_5 (Dense)                  & (None, 2)             & 66         \\
\hline
\hline
Total params: 33,649,890          &                     &  \\
Trainable params: 33,649,890      &                     &  \\
Non-trainable params: 0           &                     &   \\        
\end{tabular}
\label{tabdl}
\end{table}

From our results and these statements, we can see that we achieved almost 85\% on the first attempt, which is a significant signal of the performance of the method and model used, which can be further increased, and also that these first results are realistic and practically applicable. Finally, for the last experiment, we performed a deep learning correction. This configuration and the results are presented in section \ref{mgi}.

\begin{figure}[!ht]
\centering
\includegraphics[scale=0.6]{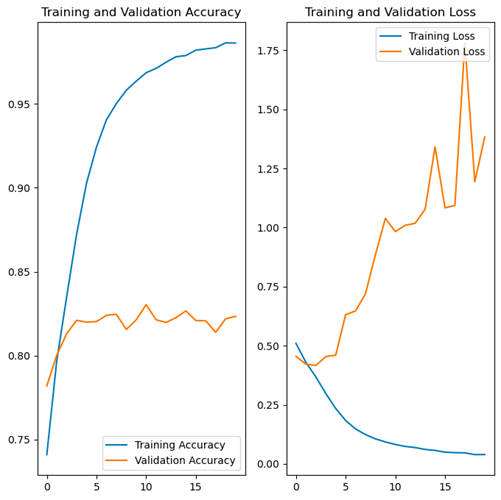}
\caption{Training and validation.}
\label{lv}
\end{figure}

When learning the neural network, it can be observed, see Figure \ref{lv}, that between epochs 5 to 15, the validation set stabilizes in terms of quality and thus, more epochs are probably not needed. In this way, we can estimate how many epochs are required to learn the network so that it is not overtrained.

The fact that there are many network parameters is because we have experimented a lot and tried different complex architectures. We compared them to each other to choose the best one with the highest validation and test data accuracy. During the experiments, we also had architectures that had significantly fewer adjustable parameters (from 2 million to 127 million). Models with more training parameters may always be needed for fractal image classification because fractal classification is much more complex than common image classification cases such as font recognition, flower classification, car recognition, etc.

During training, we monitored the evolution and accuracy of the validation data and stopped the training in time so that the model should not be overtrained. Since the model was not trained on validation data either, we can say that we tested the model on validation data of 6 427 images + on test data of 1 000 images (total 7 427 images), which is already a relevant result considering the number of training data. The test data was the same for all tested architectures and resolutions so that we could compare each other. The validation and training data for each architecture were chosen randomly from the entire set of images we selected. Of course, a more significant number of data can be tested.

\subsection{Malware and goodware fractalisation}
\label{MGF}
For experimental needs, many different visualizations of MF were created during our research, which were eventually converted into the two different graphical representations we will discuss here. These graphical ones differ from each other in terms of colour and accuracy of fractal rendering. 

These sets were designated as Set 1 and Set 2. Figure \ref{Win32.Adware.BrowseFox} shows an example of the generated MF according to the generative equations, with a fixed iteration length of the algorithm. The sample shows, and this is true for all images in this set, that all four quadrants have the same colour. This essentially follows from the constant numbers of iterations (64, usually over 40-50 no visual changes are visible in the figure) for all four quadrants. However, since this approach is a bit uniform from this point of view and a slightly different equation generates each quadrant with a different degree of generating polynomials, we decided to try a slightly modified approach. The result of this modification is shown in Fig. \ref{Win32.Adware.BrowseFoxExt}, which shows a sample MF from graphics set No. 2. In creating this fractal set, the variable number of iterations required to render each quadrant of the colour pattern was taken into account. This number of iterations corresponded to the maximum order polynomial used to generate the pattern. We found it appropriate to use this number because, in a way, it characterizes the network built by the API call of the malware in question. So this information was then reflected in the images.

As one can see from the two pictures, the result is very different. The fractals from set 1 are colour-uniform images, on the other hand, with very high accuracy in rendering the individual parts of the fractal. On the other hand, the fractals generated by the modified approach are much richer in colour. Still, due to the variable length of the iterations for rendering according to the TEA algorithm, the fractals are often drawn with less detail.

Experimental validation showed that the classification performance only increased by 0.5\% for the Set 2 series. This implies that this change, although visually visible, has no particular impact on classification quality. Therefore, only the Set 1 image was used for further experiments because of the higher precision of generated images.

\subsection{Mutation?}
\label{mutmal}
As part of the preparation for our experiments, the generated images were checked, and some images were identical. 
On closer examination, we found that if we consider the virus representation as a graph and not as a linear string of API sequences, then we get very similar structures that differ only usually in the last few API libraries, that are reused many times in a row. In other words, linear API call strings contain a small amount of the same and then differ in large parts just by repeating two or a few different APIs. In graph terms, this repetition simplifies to just a few vertices in the graph, and the malware's similarity becomes more apparent using our view. 

\begin{figure}[!ht]
\begin{minipage}[b]{0.5\linewidth}
\centering
\includegraphics[scale=0.7]{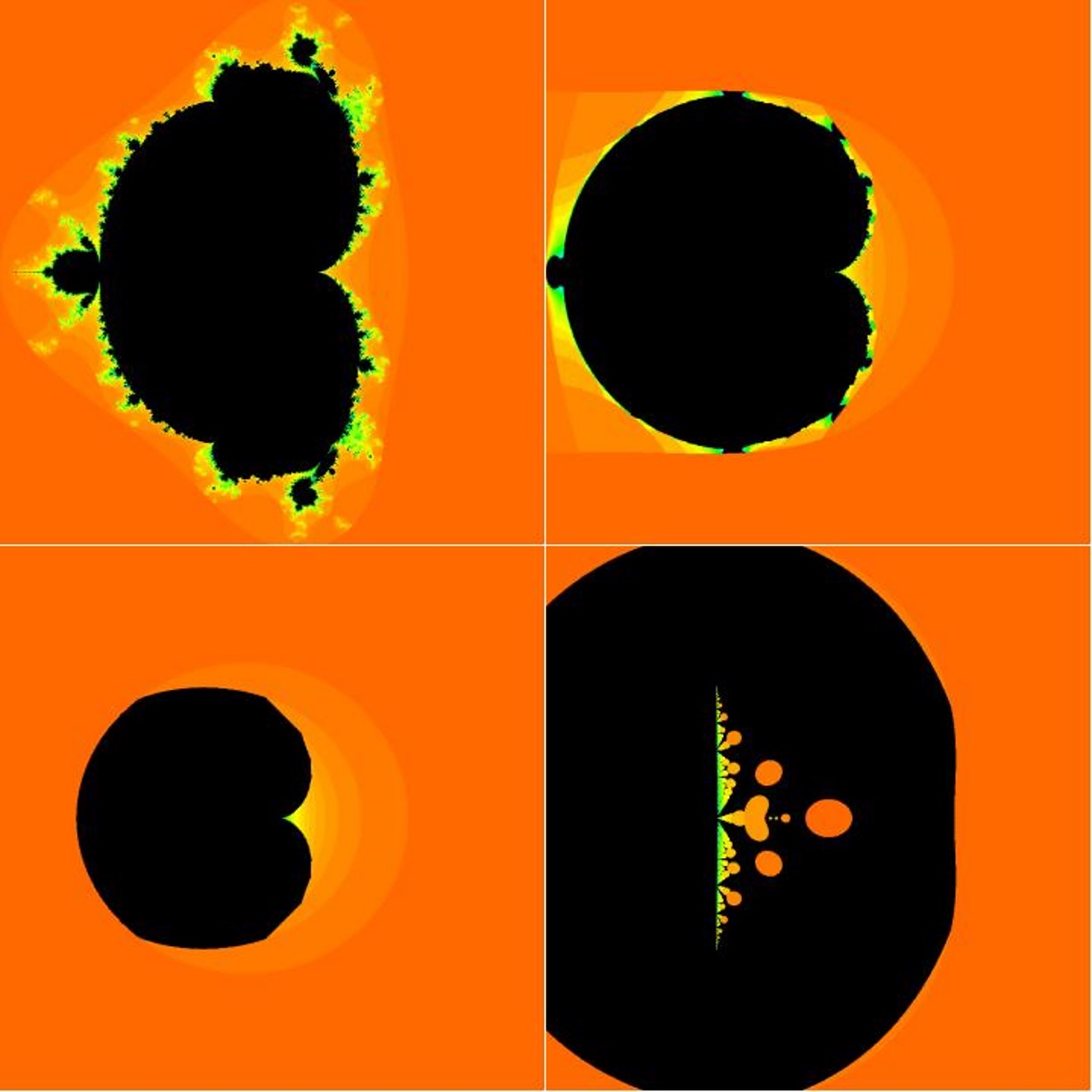}
\caption{One of the many instances of an image that has been found in a different malware. However, as can be seen in the image \ref{podobnost1}, \ref{podobnost2}, \ref{podobnost3}, the behavioural structure of the malware is practically the same.}
\label{stejnyobrazek}
\end{minipage}
\hspace{0.5cm}
\begin{minipage}[b]{0.5\linewidth}
\centering
\includegraphics[scale=0.7]{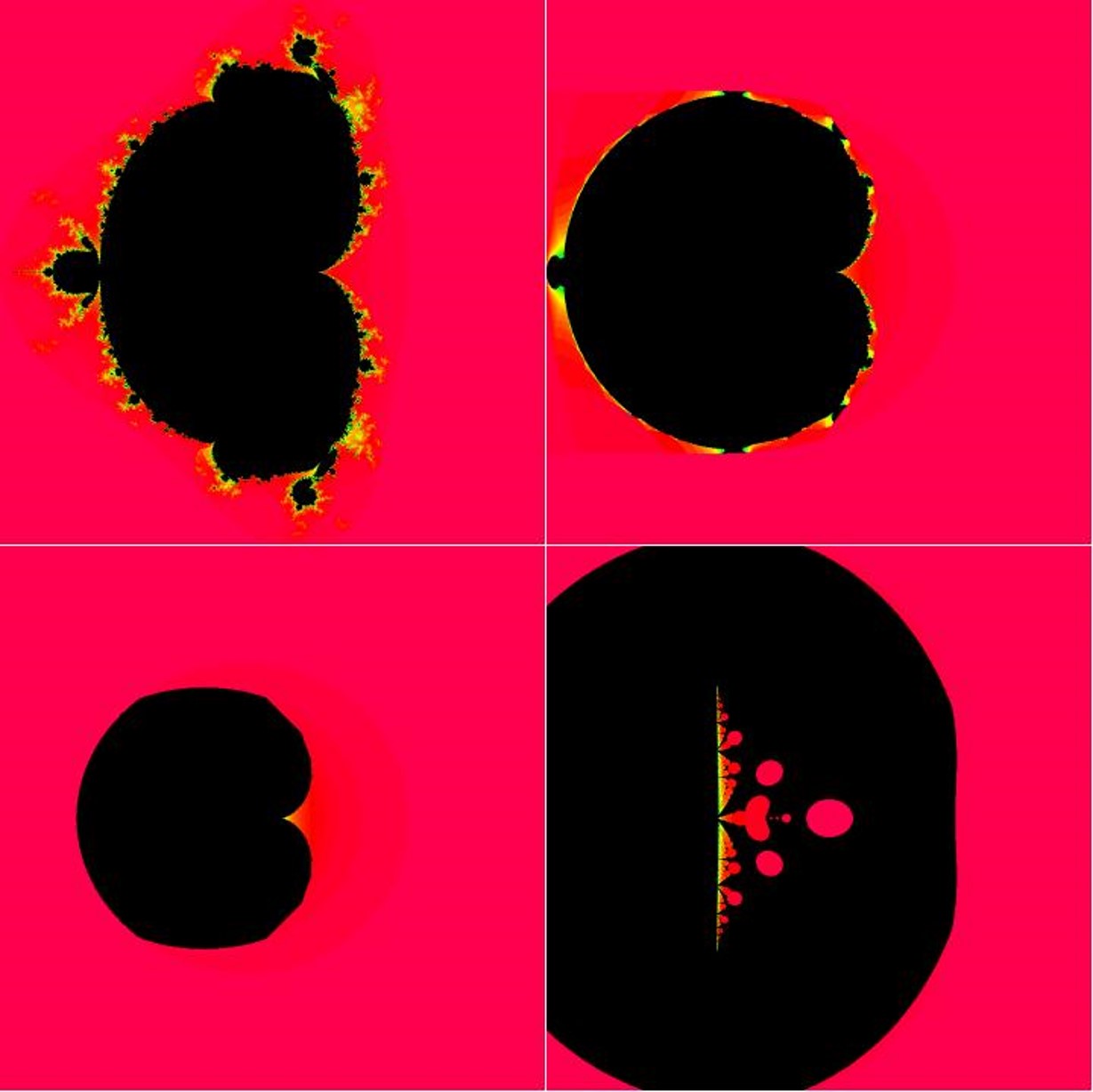}
\caption{Another case of a display that was the same but with a different colored background. A color-coded malware mutation (see Fig. \ref{podobnost3})?\\ \\}
\label{jinyobrazek}
\end{minipage}
\end{figure}

\begin{figure}[!ht]
\centering
\includegraphics[scale=0.5]{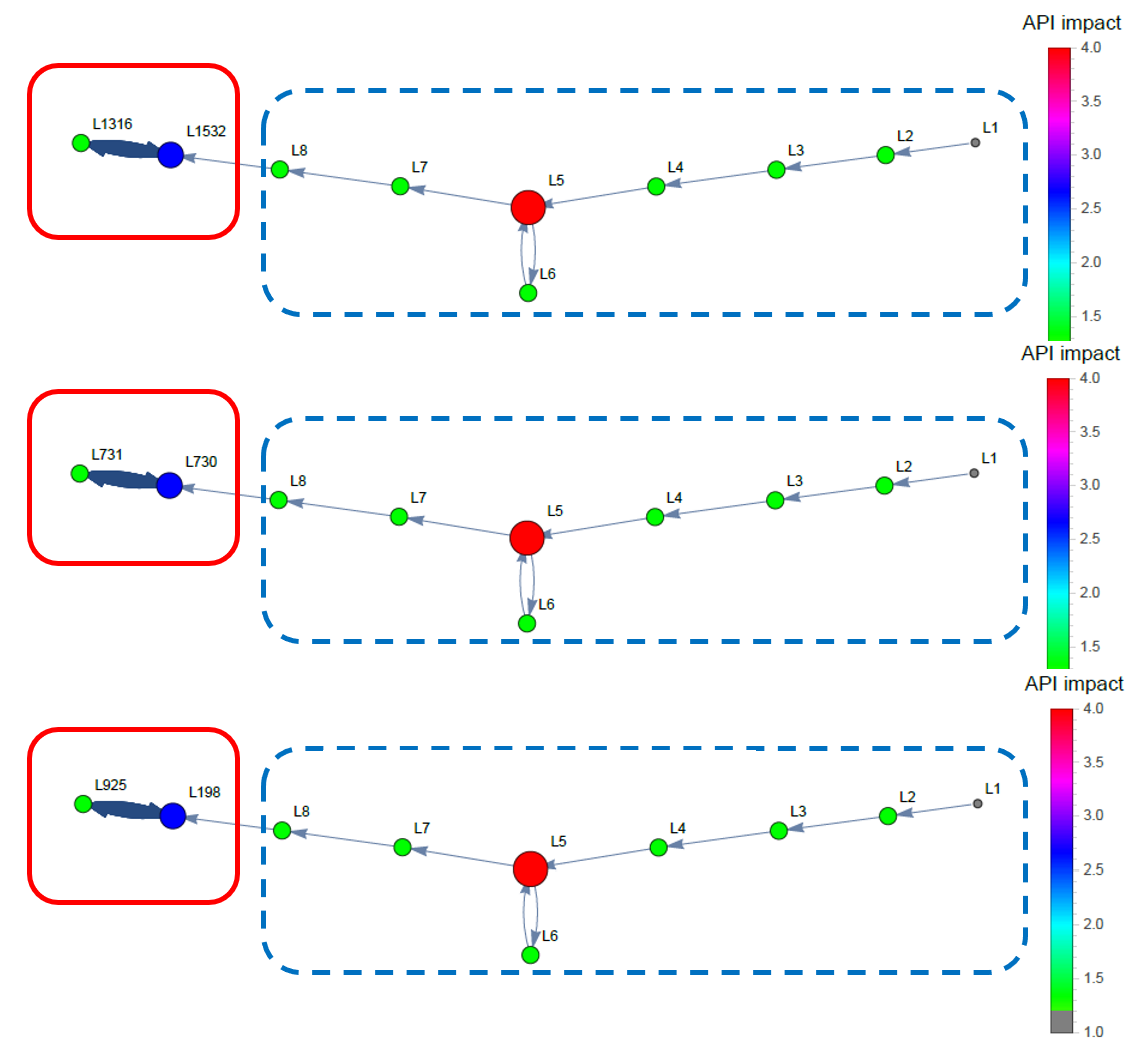}
\caption{Differences in samples (the red box).}
\label{podobnost1}
\end{figure}

\begin{figure}[!ht]
\centering
\includegraphics[scale=0.5]{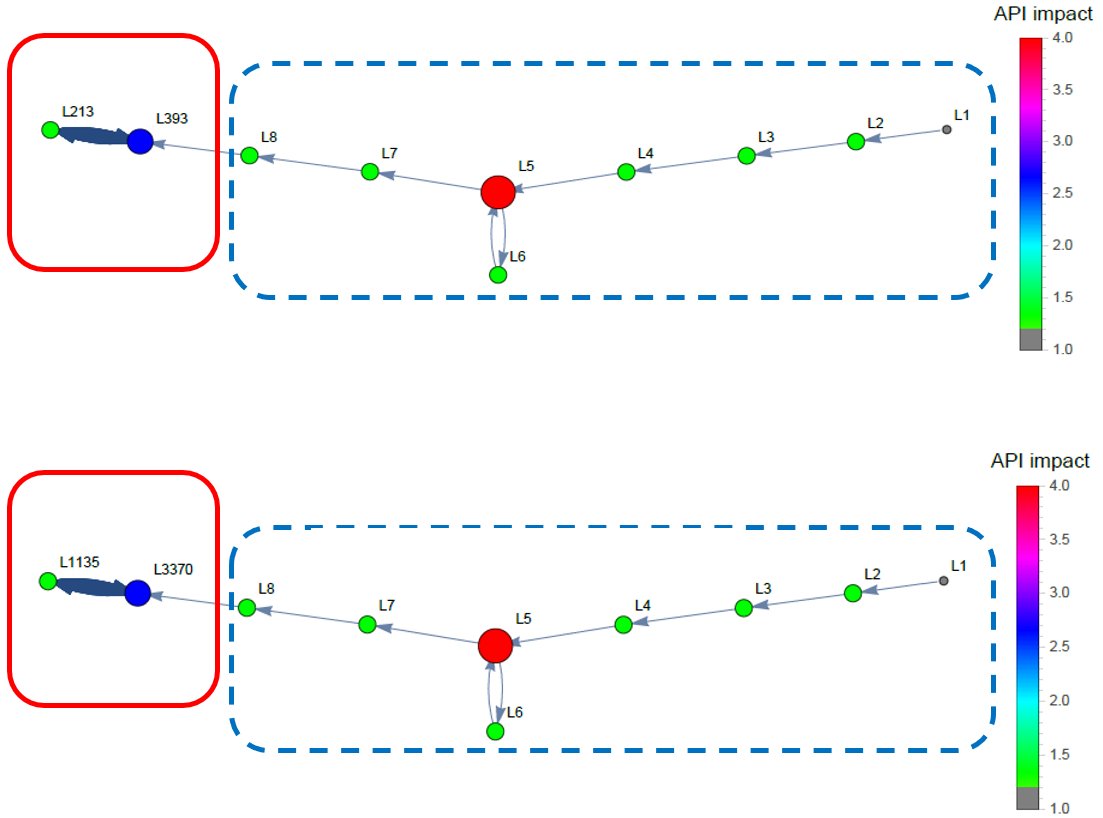}
\caption{2 $\times$ Differences in samples (the red box)?}
\label{podobnost2}
\end{figure}

\begin{figure}[!ht]
\centering
\includegraphics[scale=0.35]{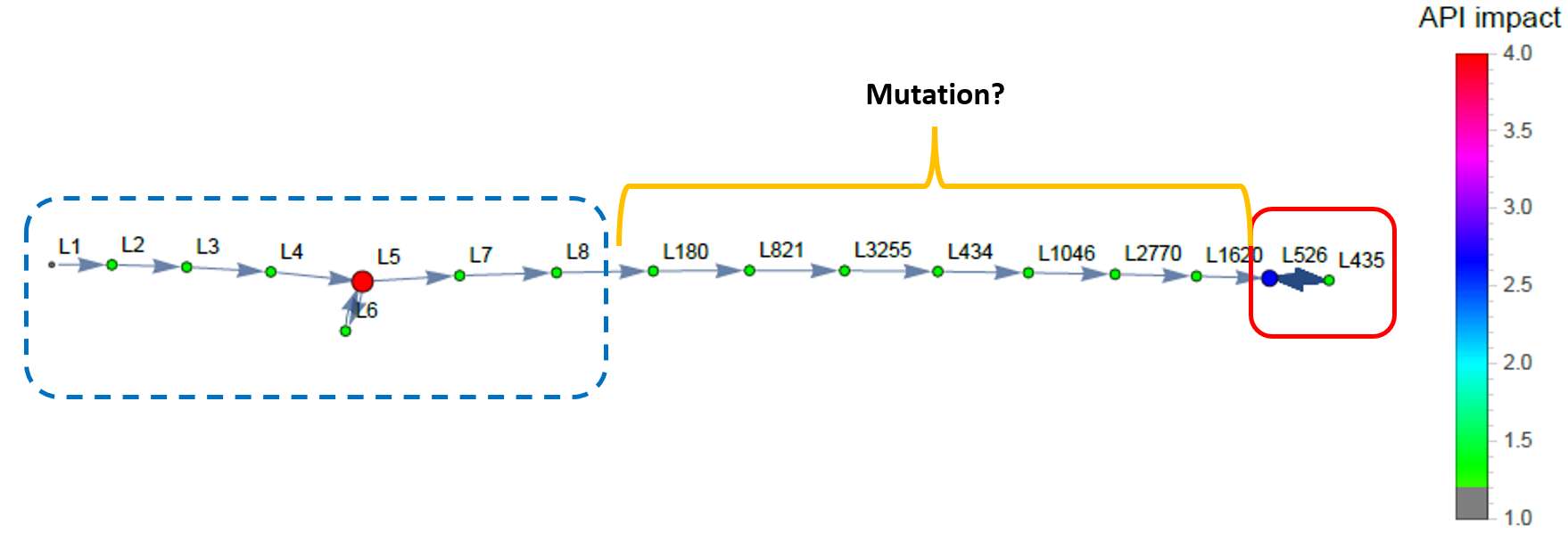}
\caption{Different sample - another case, in this case apparently a mutation, of the same malware, see Fig. \ref{jinyobrazek}. Compare with Figs. \ref{podobnost1} and \ref{podobnost2}.}
\label{podobnost3}
\end{figure}

From the examples in the figures (Fig. \ref{stejnyobrazek}. \ref{jinyobrazek}), \ref{podobnost1},\ref{podobnost2} a \ref{podobnost3} it is pretty clear that these different viruses are de-facto only modifications of themselves. At least based on APIs sequences analysis.
Also, suppose there is a change in the behavioural structure (API sequence) of a given virus in a minor way. In that case, our imaging method captures this in the form of, for example, a colour change, as seen in the images Fig. \ref{jinyobrazek} a \ref{podobnost3}. Thus, our imaging method can clearly detect the similarity of viruses (a bit different APIs sequences). Our approach also allowed the identification of more significant "mutations" in the form of colour background imaging.

\section{Results} 
\label{mgi}
The aim of this identification was to distinguish goodware from malware. The classification was based on 4 malware images (each of size 512x512) combined into one, see Fig. \ref{Cryptowall} - \ref{Petya}. 
The experimental dataset contained a total of 65,270 images, respectively 32,097 malware and 33,173 goodware. 

This dataset was divided into 3 groups in the experiment: 
\begin{itemize}
    \item Training group 57,843 images (malware and goodware)
    \item Validation group 6,427 images (malware and goodware)
    \item Test group 1,000 images (malware and goodware)
\end{itemize}

\begin{figure}[!ht]
\centering
\includegraphics[scale=0.45]{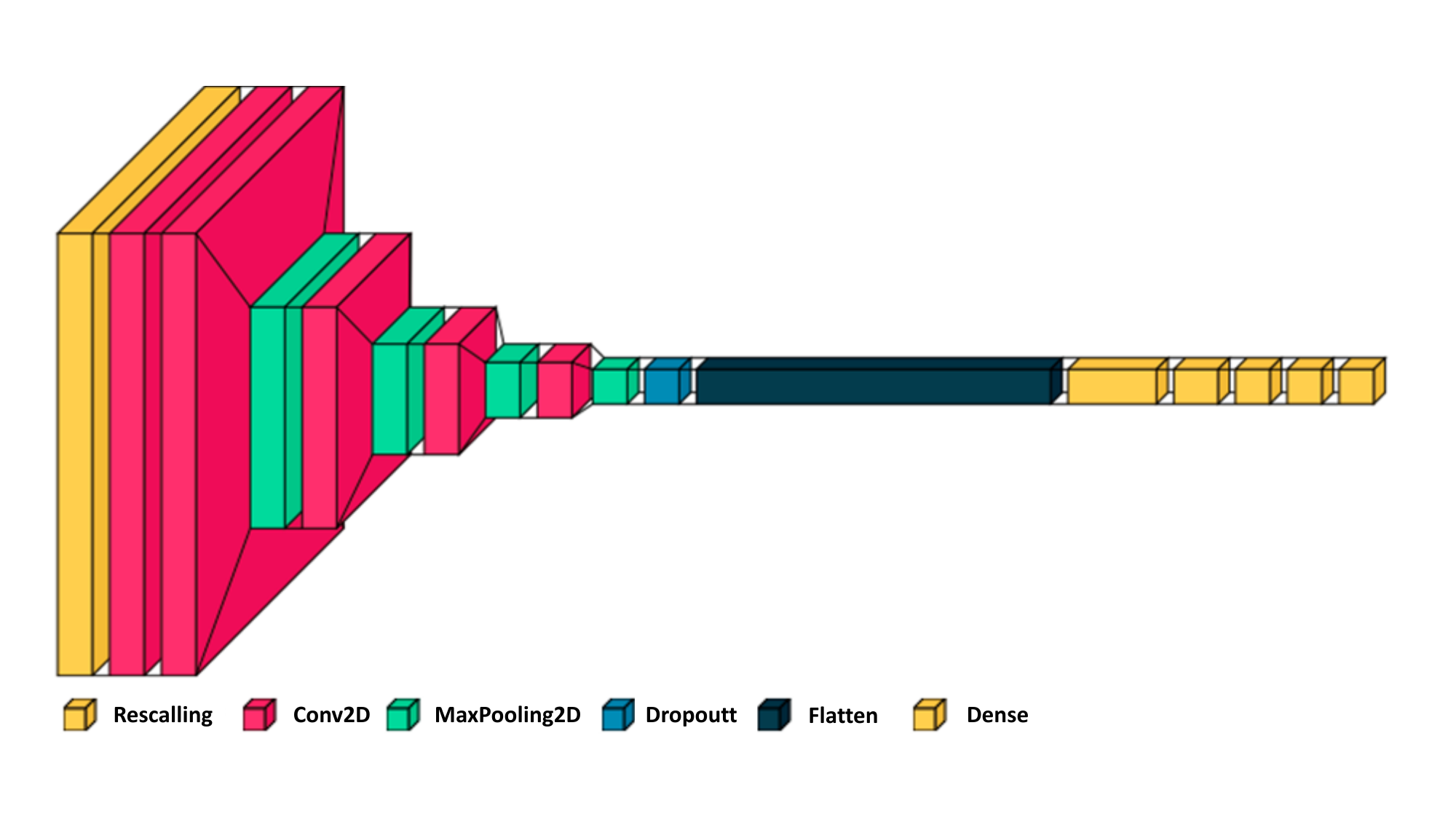}
\caption{Deep learning structure}
\label{dl2}
\end{figure}

\begin{figure}[!ht]
\centering
\includegraphics[scale=0.6]{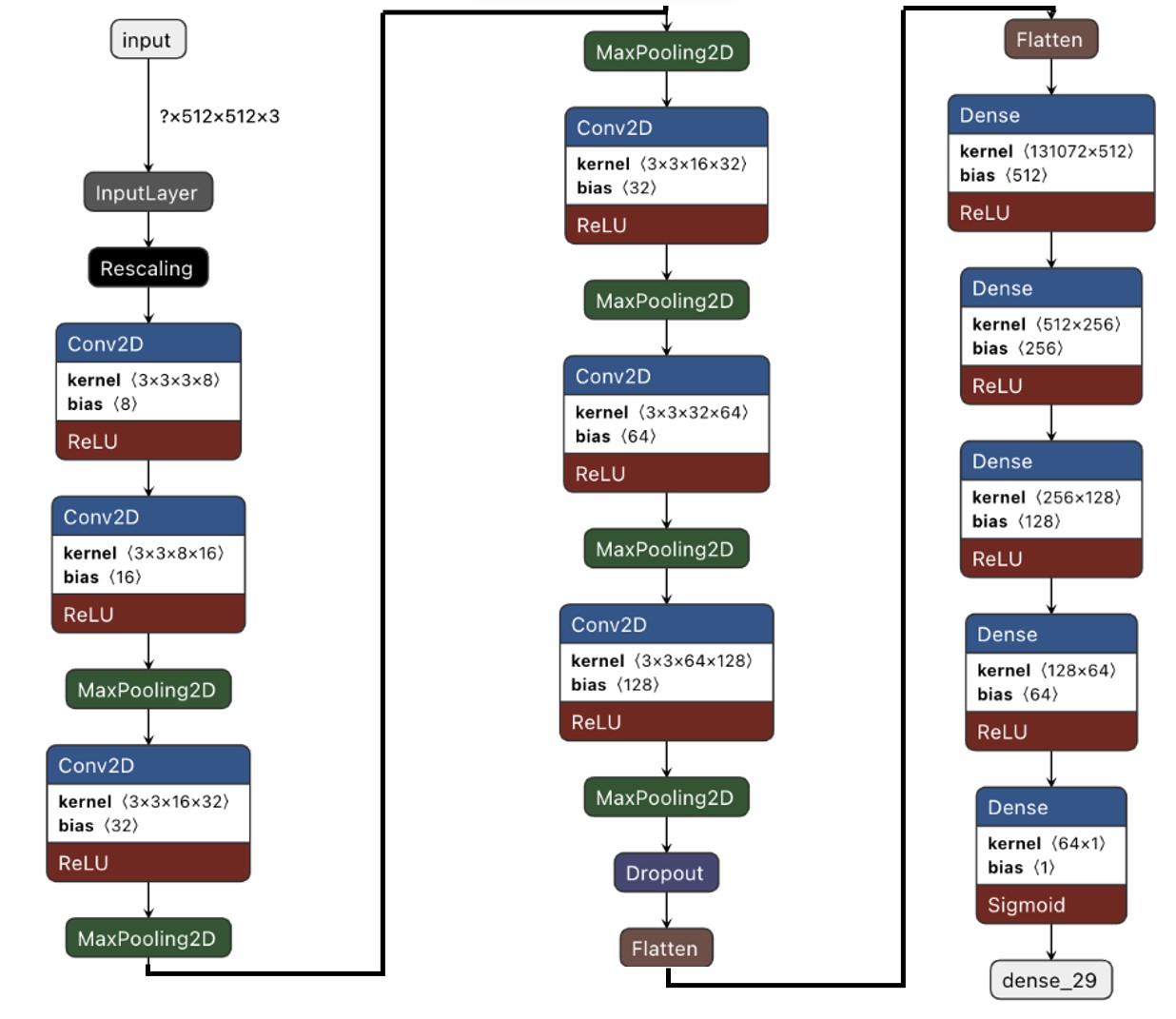}
\caption{Deep learning structure}
\label{dl3}
\end{figure}

A convolutional neural network (CNN) was chosen to classify this dataset. This network is implemented in the Python programming language using the Tensorflow 2.0 library. 
Before working with the convolutional network, preprocessing/normalization of the data is done. The dataset contains coloured images; hence, it is better to normalize the RGB values from the 0 - 255 range to the 0-1 range. For experiments, we have created a simple convolutional neural network model, see Fig. \ref{dl2}, \ref{dl3} and Tab. \ref{tabdl2}. 
The resulting modified mesh has a first image rescaling layer, then two convolutional layers, followed by a MaxPooling2D layer, and then three convolutional layers always paired with a MaxPooling2D layer. These layers are followed by a \textit{Dropout} and \textit{Flatten} layer, followed by a fully connected neural network. This fully connected network contains four layers + 1 output neuron with Sigmoid activation. This network uses the optimizer \textit{Adam} and the loss function \textit{BinaryCrossEntrophy} in training. We can see the model summary in Tab. \ref{tabdl2} for more details.

\begin{table}[!ht]
\centering
\caption{Used improved model, see also Fig. \ref{dl2} and \ref{dl3}. For results see Tab. \ref{60000} and \ref{tptn}.}
\small
\begin{tabular}{lll}
\hline
Layer (type)                      & Output Shape          & Param \#   \\
\hline
rescaling\_1 (Rescaling)     	& (None, 512, 512, 3)       & 0\\
conv2d (Conv2D)             	& (None, 512, 512, 8)       & 224\\
conv2d\_1 (Conv2D)           	& (None, 512, 512, 16)      & 1168\\  
max\_pooling2d (MaxPooling 2D)    & (None, 256, 256, 16)      & 0\\ 
conv2d\_2 (Conv2D)           	& (None, 256, 256, 32)      & 4640\\  
max\_pooling2d\_1 (MaxPooling 2D) & (None, 128, 128, 32)      & 0\\ 
conv2d\_3 (Conv2D)               & (None, 128, 128, 64)      & 18496\\ 
max\_pooling2d\_2 (MaxPooling 2D) & (None, 64, 64, 64)        & 0\\ 
conv2d\_4 (Conv2D)               & (None, 64, 64, 128)       & 73856\\
max\_pooling2d\_3 (MaxPooling 2D) & (None, 32, 32, 128)       & 0\\ 
dropout (Dropout)               & (None, 32, 32, 128)       & 0	\\         
flatten (Flatten)               & (None, 131072)            & 0	\\         
dense (Dense)                   & (None, 512)               & 67109376\\  
dense\_1 (Dense)                 & (None, 256)               & 131328\\    
dense\_2 (Dense)                 & (None, 128)               & 32896\\     
dense\_3 (Dense)                 & (None, 64)                & 8256\\      
dense\_4 (Dense)                 & (None, 1)                 & 65\\        
\hline
\hline
Total params: 67 380 305        &			                &	\\
Trainable params: 67 380 305    &			                &	\\
Non-trainable params: 0         &					        &	\\
\end{tabular}
\label{tabdl2}
\end{table}

The resulting model had an accuracy of 98.62\% over the training set and 83.21\% over the validation set. Finally, the model was tested on the test set. The result on the test set achieved an accuracy of 85.7\%. 
Unlike some previous research publications, where the numbers of malware and goodware were heavily skewed in favour of malware, our image set was very rich and balanced. That is, the number of malware and goodware was almost equal. Furthermore, the images were used at a resolution of $512\times512$ and the same distribution as mentioned at the beginning of the section.


\begin{table}[!ht]
    \centering
    \begin{threeparttable}
    \caption{Results for experiment 2.}
    \label{60000}
    \begin{tabular}{l|l|l|l|l}
        ~ & Total & Training & Validation & Test  \\ \hline
        \hline
        Total & 65 270 & 57 843 & 6 427 & 1 000  \\ \hline
        Malware & 32 097 & 28 437 & 3 660 & 500  \\ \hline
        Goodware & 33 173 & 29 406 & 3 767 & 500  \\ \hline
       \hline
        Precision & - & 98.62 \% & 83.21 \% & \textbf{85.7} \% 
    \end{tabular}
        \begin{tablenotes}
      \small
      \item Specific test results:\\
437 correctly classified Malware, 63 incorrectly (87.4\%).\\
420 correctly classified Goodware, 80 incorrectly (84.0\%).
    \end{tablenotes}
  \end{threeparttable}
\end{table}

\begin{table}[!ht]
    \centering
    \caption{F1 et all for results from experiment 2.}
    \label{tptn}
    \begin{tabular}{l|l|l|l|l|l}
        ~ & Total & TP & TN & FP & FN  \\ \hline \hline
        Malware & 500 & 437 & - & 80 & -  \\ \hline
        Goodware & 500 & - & 420 & - & 63  \\ \hline \hline
        Accuracy & 0.857 & - & - & - & -  \\ \hline
        Precision & 0.845 & - & - & - & -  \\ \hline
        Recall & 0.874 & - & - & - &   \\ \hline \hline
        F1 & 0.865 & - & - & - & - 
    \end{tabular}
\end{table}

For the results of these experiments, see Tab. \ref{60000} and \ref{tptn} have confirmed the preliminary results obtained in Tab. \ref{tbl}, and with a few experiments with different configurations of deep learning networks, these are the best results so far. However, as mentioned previously, these are not the final results. At this point, some ways in which the classification could be modified to improve the method's performance are straightforward. We also leave these possibilities open to the readers, who are hereby invited to try their ideas to improve the performance of the MF classification.

Comparing our results, it is pretty clear that the currently published results of other researchers are better, but it should be noted that our dataset was extensive, balanced and not, to put it bluntly, a few thousand malware samples and a few hundred goodware samples (in the extreme case). 
Of all the research reports that can be found in scientific databases, one thing is clear; to date, there is no single, unified, truly data-balanced, highly comprehensive database of relevant samples and records that the scientific community can use as a test benchmark for classification methods in malware and goodware. Such benchmarks already exist in other fields, such as test functions for optimization methods based on swarm intelligence or evolutionary techniques. This is another topic to think about - creating such a database.

Moreover, the facts mentioned above point to the high potential of the method in increasing classification accuracy, but there were other goals of our research. The main goal is to show the possibility of how malware and goodware can be visualized in a completely different way and thus open a research discussion on this topic.


\subsection{Open questions and new research directions} 
The research we report brings a lot of exciting research questions to our attention - for example, malware visualization. Our contribution is that computer malware images are not cubic grey structures but de facto fractal patterns with their regularities and properties. So, for example, the question is how the fractal dimension of such images is related to the properties of malware. Is the fractal dimension higher when the complexity of the computer virus behaviour is higher? Can fractal dimension be used to categorize malware into groups and families based on behavioural similarity? Can the escape trajectory speed when plotting a fractal pattern be used to more accurately represent malware characteristics (which, incidentally, was demonstrated in one of two experiments with 65,270 images)? Other topics for discussion may be as follows.

\textbf{Different visualization modifications}. In our case, we have done several different experiments with visualization methods, and the one we have chosen here is considered representative. However, this does not mean that there are no other, possibly better, modifications of our image-generating methods within fractal geometry. The possibilities for constructing fractal patterns are quite rich, and at least there are still many unanswered questions.

\textbf{Similarity of recorded behavior}. As discussed above, we found the same images corresponding to different APIs when working with the data. However, upon closer analysis, we found that they are likely the same malware with slight modifications (Sec. \ref{mutmal}), which was revealed through our conversion to a graph and subsequent display. Thus, the question is whether our proposed methods can refine the virus databases and their classification and thus make them more accurate and better organized in the classification framework.

\textbf{Static and dynamic analysis}. The research and results we report are only related to the dynamic analysis of computer malware behaviour. Thus, the images we present reflect the behaviour of the malware itself. However, it is known that there is a static analysis in addition to the dynamic analysis, which is the source of the grayscale images already mentioned. The question is, therefore, whether our method could be applied to static analysis of binaries as well, with the idea that it could replace existing grayscale images with fractal images similar to those presented in our work, and thus, of course, bring many interesting questions to further research on static analysis in malware representation.

\textbf{Modification of TEA algorithm.} In our research, we used the standard TEA algorithm as reported in the literature \cite{barnsley2014fractals}. This resulted in beautiful fractal patterns that passed the classification test. The question is whether this algorithm can be modified to emphasize better the network structure, the importance of individual API calls, and possibly other malware attributes. This opens the topic of research into modifying the imaging algorithm itself.

\textbf{Better deep learning model.} All the results were obtained using more or less standard deep learning techniques. However, this does not mean that it is impossible to find better configurations of the networks used or modify them to create our own that will have better performance. This, even with the possible application of the previous points, may again be a fascinating topic for further research in the visual classification of malware.

\textbf{Better dataset.} The dataset we used, although very comprehensive, was only a sliver of the total amount of data that is theoretically available. The question is how the method's performance would have changed if a much larger dataset of records had been used so that all relevant malware samples were represented evenly and not dominated by a few families. Especially in the case of classifying malware into families.

\textbf{Mutation of malware and its gradual evolution.} As mentioned earlier in the paper, our method can be used to capture even small changes in the malware structure in the form of visualization, both in the form of colours and possibly subtle changes in the fractal form itself. The question still needs to be answered is whether the gradual mutation of malware can be visualized using our representation. The answer is probably positive (see section \ref{mutmal}). However, this topic also deserves further research, and it would be fascinating to see the malware mutation in the form of a fractal representation evolving over time. Perhaps this could also predict the future form of the mutation in visual form.

\textbf{Reverse identification of malware and goodware.} So far, we have identified malware and goodware using deep learning methods in the images generated by the classical TEA algorithm. In conjunction with whether TEA can be modified to better capture the characteristics of malware and goodware, the question arises whether reverse identification can be performed just by looking at the images. This would undoubtedly be an added value of this method in terms of classification. So again, this is another open topic for research.

There are undoubtedly many other topics that could be discussed. However, it is beyond the scope of this article, and we intend to develop and publish some of these other topics as a continuation of this research and offer these ideas for readers to consider.

\section{Conclusion and Future Work}  
\label{chapter:famCONCL}
In this paper, new methods for visualizing computer malware, precisely their dynamic behaviour, were discussed. Fractal geometry was used for this purpose, and the specific approach presented here was used to show how different types of malware can be visualized in the form of fractal patterns—the reason why fractal geometry, in particular, is quite apparent. Not only does this style of display provide visually appealing images, but the structure of the iterative polynomial equation is reflected in the structure of the image by its geometric properties, including colours. It is thus a conversion, figuratively speaking, of \textit{from the space of equations to the space of fractals} (it is not a mathematically defined space, but an analogy). In fractal geometry, there is now a comprehensive mathematical apparatus containing a description, analysis, and calculation of fractal dimensions or work with hyperfractals. The applicability of this area of geometry is nowadays extensive indeed. Beyond its visual appeal, this representation is not an end in itself but represents a new area for research in malware analysis.

Using malware fractal representation, we used a deep learning network to classify malware and goodware. As we mention in the paper for our experiments, a total of over 130,540 images were used in the experiments (resolution dependence analysis with different graphical representations of Fig. \ref{Win32.Adware.BrowseFox} - \ref{Win32.Adware.BrowseFoxExt}, ), to verify the possibility of using fractal geometry. This was done on randomly selected samples drawn from a database of 7,712,797 samples (4,833,893 of goodware and 2,878,904 of malware) provided by ESET\footnote{https://www.eset.com/}. In these experiments, we investigated the dependence of classification quality on image resolution; based on this, we chose a single experiment with 65,270 malware and goodware samples. All the results are clearly displayed in the corresponding tables.

Of course, the nature of the method and what fractal geometry itself has to offer raises many exciting research areas and questions, which we also mentioned in the previous section. These new areas of research deserve attention and hold the potential for new and novel insights into malware classification, analysis and study.

\section{Acknowledgment}
Our sincere and profound thanks to ESET s.r.o., which has supported us selflessly in our research for the last three years with data and consultancy. This work and research would never have been possible without this cooperation and support. For this, \textbf{great appreciation and our gratitude} is due to this company.

The following grants are acknowledged for the financial support provided for this research: grant of SGS No. SP2022/22, VSB-Technical University of Ostrava, Czech Republic.

\section{Appendix - gallery of selected malware fractals}
\label{gallery}
A few selected figures for an inspiration is released here.

\begin{figure}[!ht]
\setkeys{Gin}{width=0.24\linewidth}
\subfloat{\includegraphics{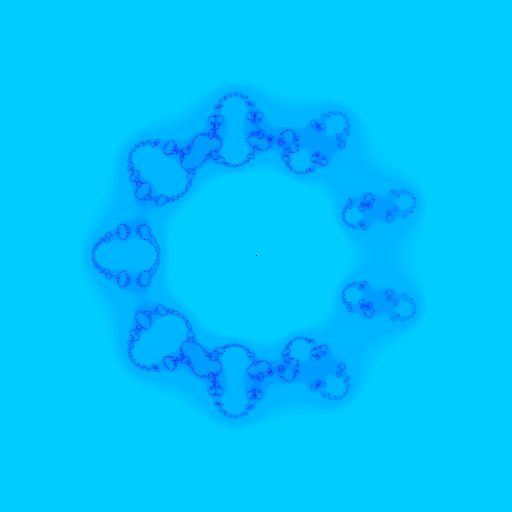}}\hfill
\subfloat{\includegraphics{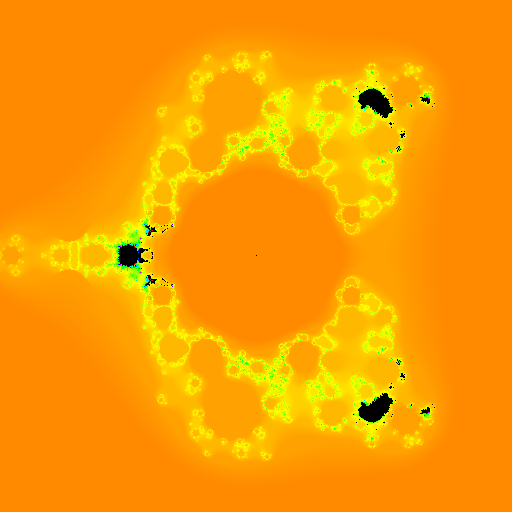}}\hfill
\subfloat{\includegraphics{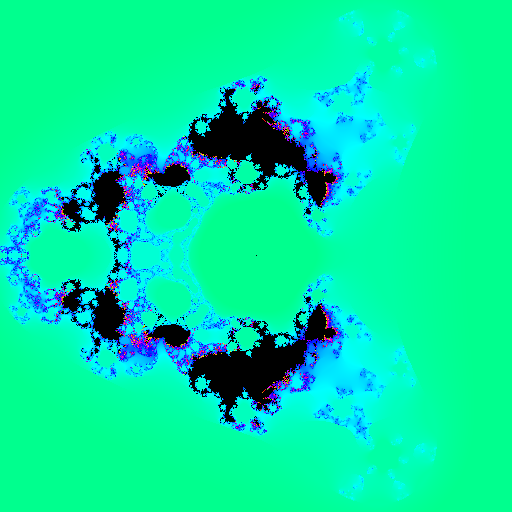}}\hfill
\subfloat{\includegraphics{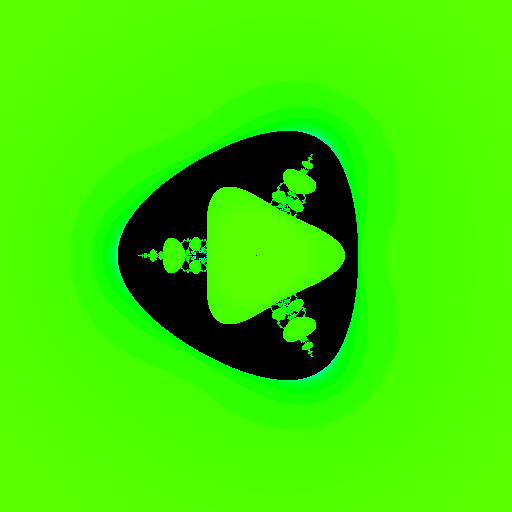}}\\
\subfloat{\includegraphics{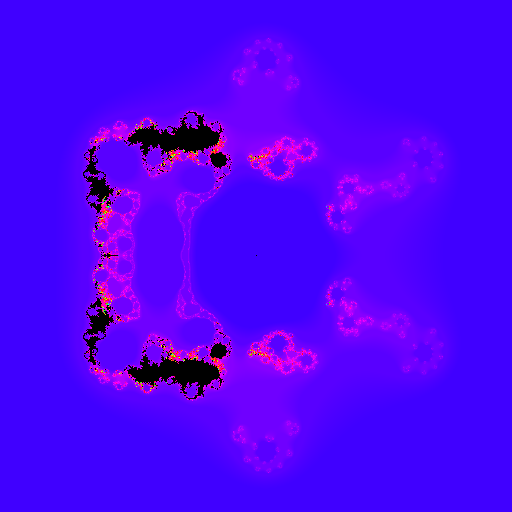}}\hfill
\subfloat{\includegraphics{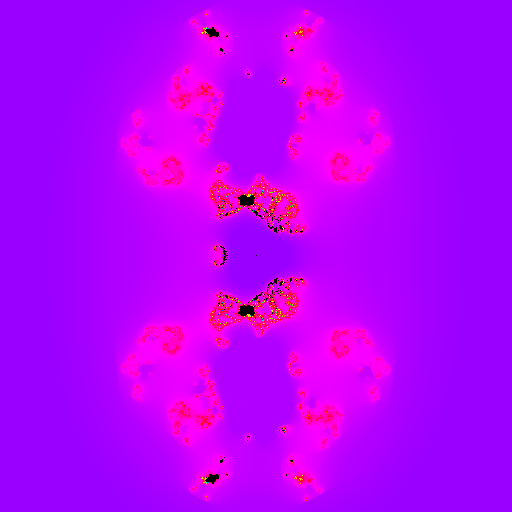}}\hfill
\subfloat{\includegraphics{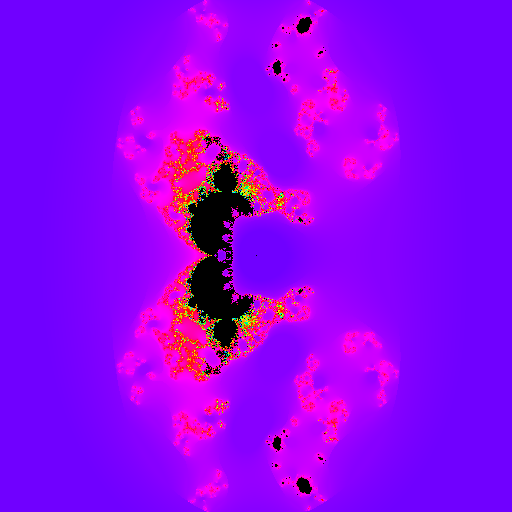}}\hfill
\subfloat{\includegraphics{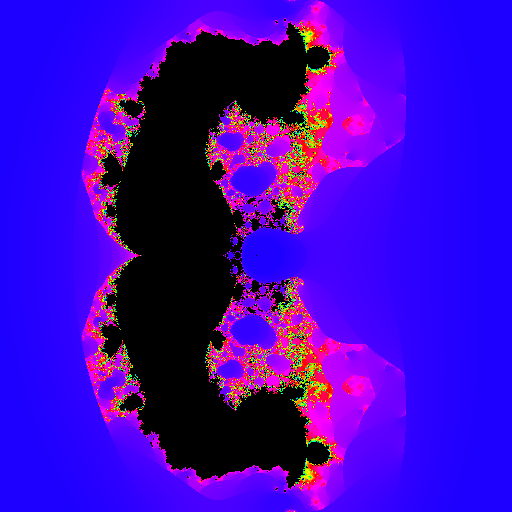}}\\
\caption{Malware samples. 
}
\end{figure}

\begin{figure}[!ht]
\setkeys{Gin}{width=0.24\linewidth}
\subfloat{\includegraphics{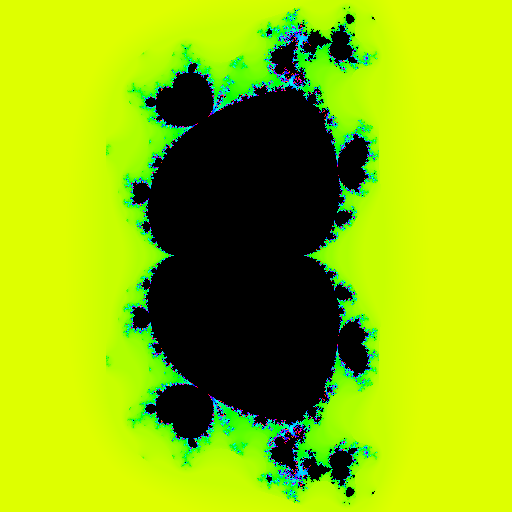}}\hfill
\subfloat{\includegraphics{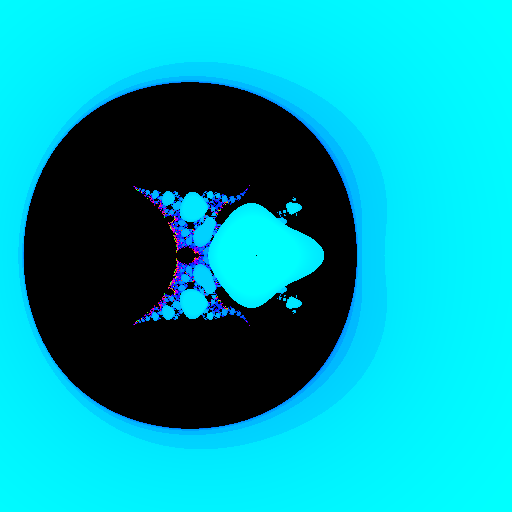}}\hfill
\subfloat{\includegraphics{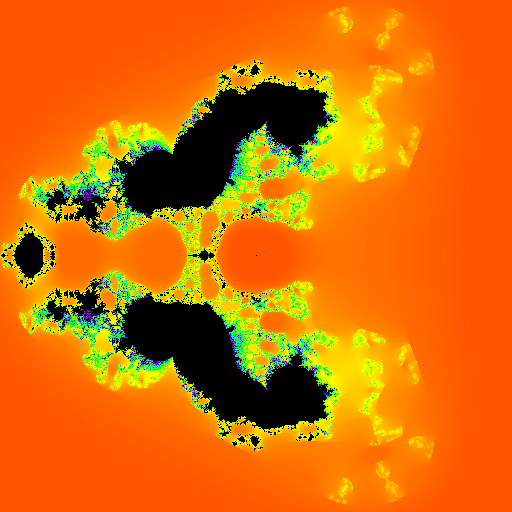}}\hfill
\subfloat{\includegraphics{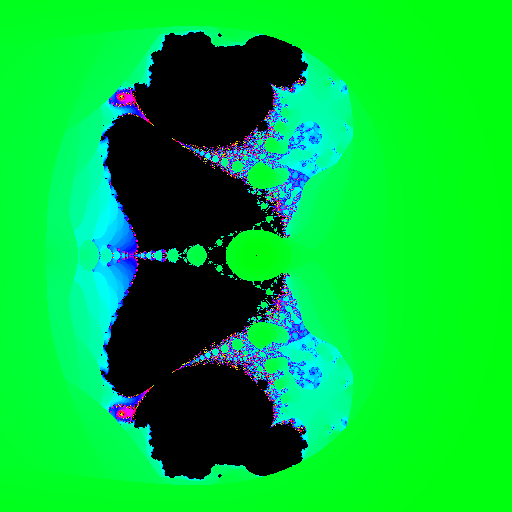}}\\
\subfloat{\includegraphics{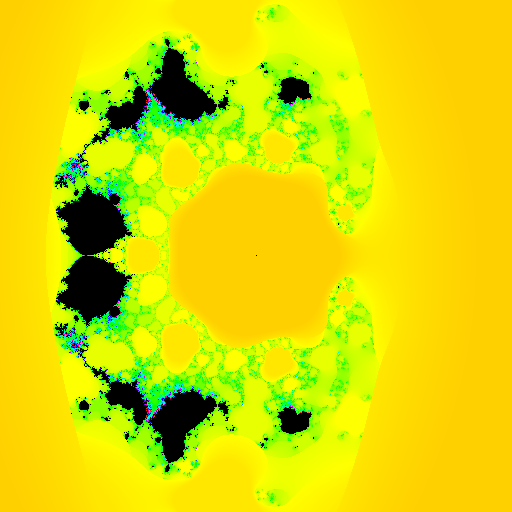}}\hfill
\subfloat{\includegraphics{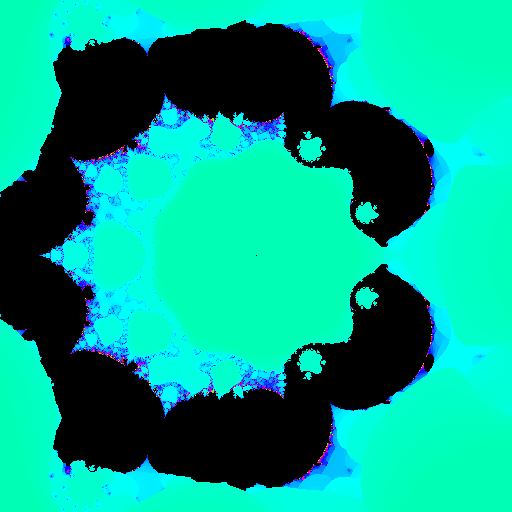}}\hfill
\subfloat{\includegraphics{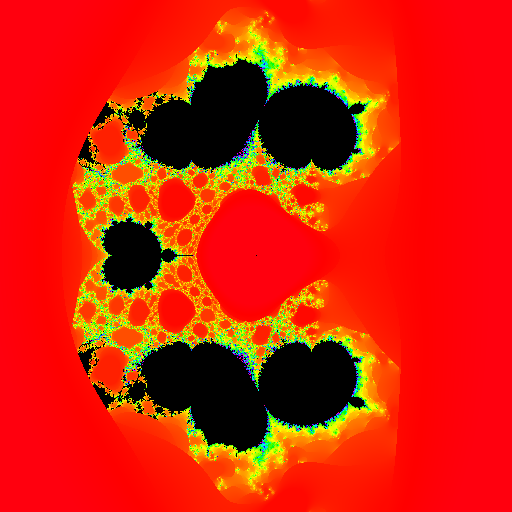}}\hfill
\subfloat{\includegraphics{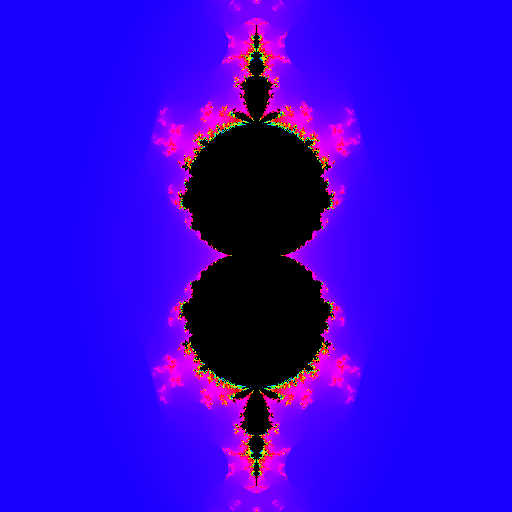}}
\caption{Malware samples. 
}
\end{figure}

\begin{figure}[!ht]
\setkeys{Gin}{width=0.24\linewidth}
\subfloat{\includegraphics{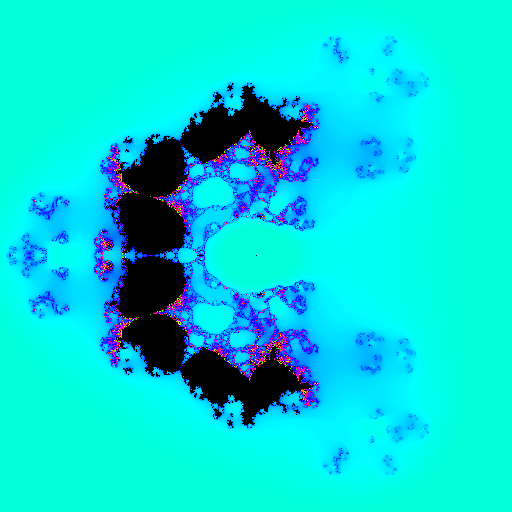}}\hfill
\subfloat{\includegraphics{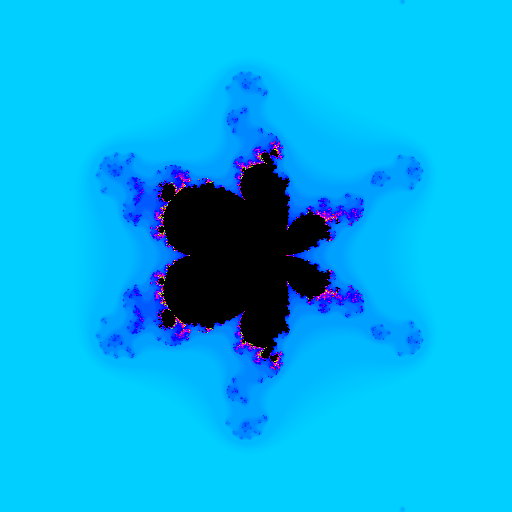}}\hfill
\subfloat{\includegraphics{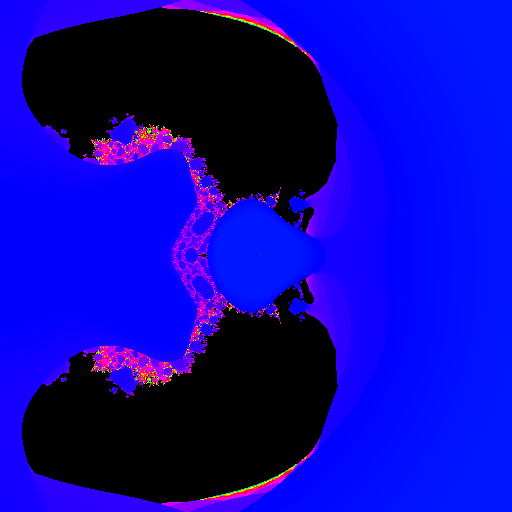}}\hfill
\subfloat{\includegraphics{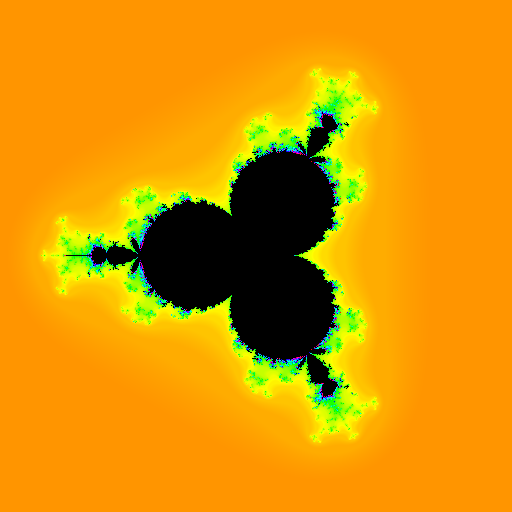}}\\
\subfloat{\includegraphics{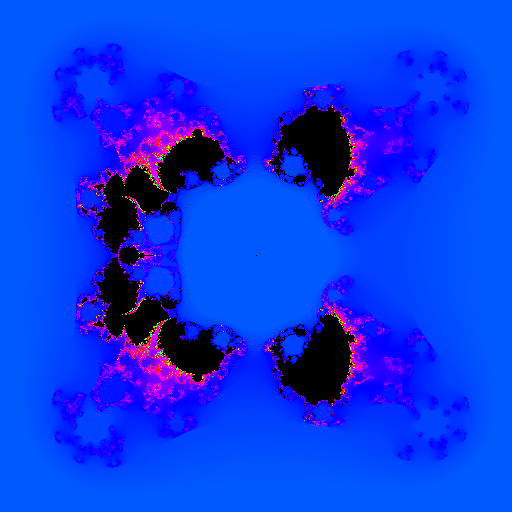}}\hfill
\subfloat{\includegraphics{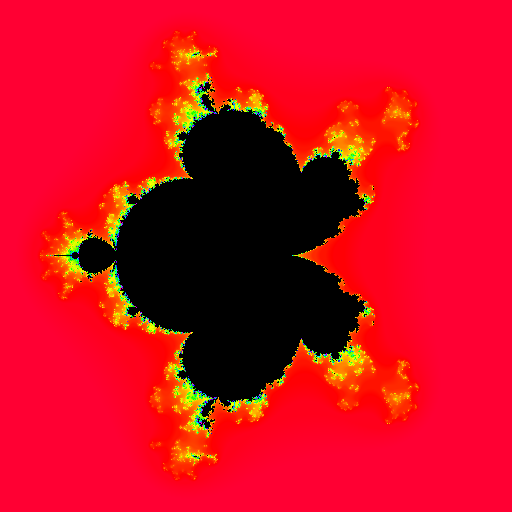}}\hfill
\subfloat{\includegraphics{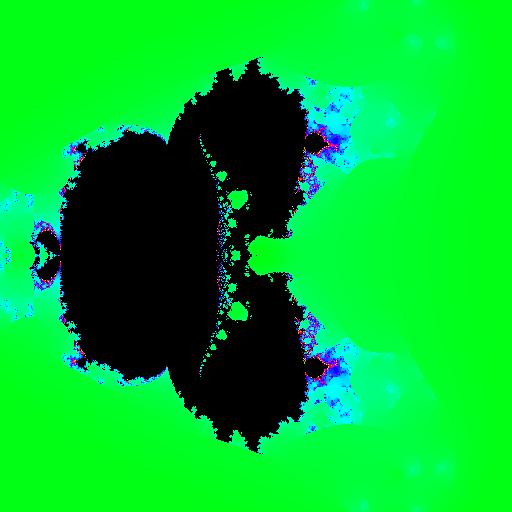}}\hfill
\subfloat{\includegraphics{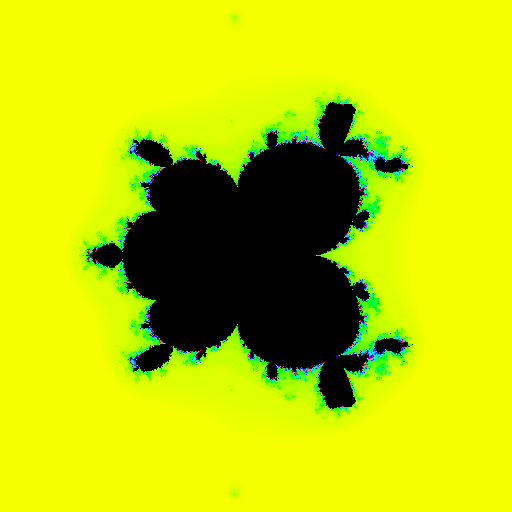}}
\caption{Malware samples. 
}
\end{figure}

\begin{figure}[!ht]
\setkeys{Gin}{width=0.24\linewidth}
\subfloat{\includegraphics{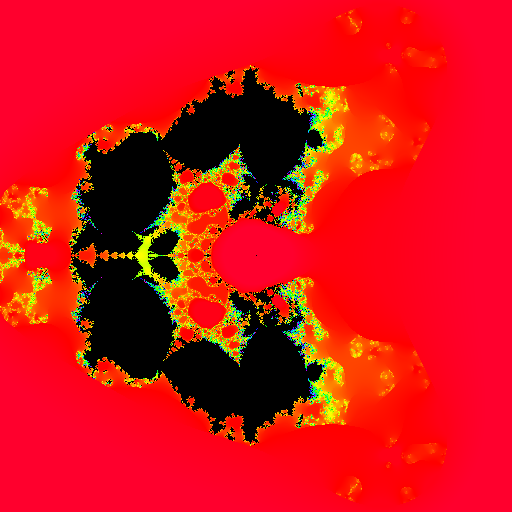}}\hfill
\subfloat{\includegraphics{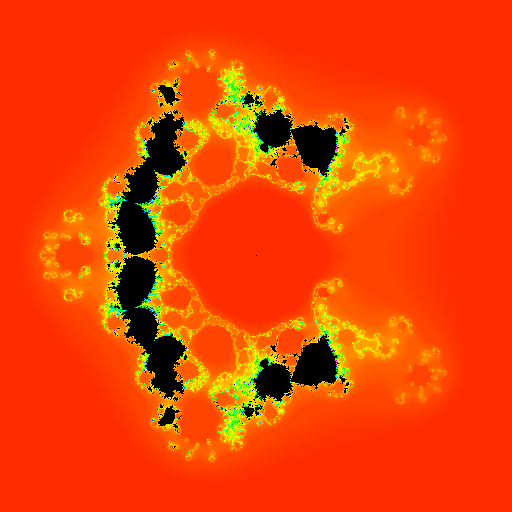}}\hfill
\subfloat{\includegraphics{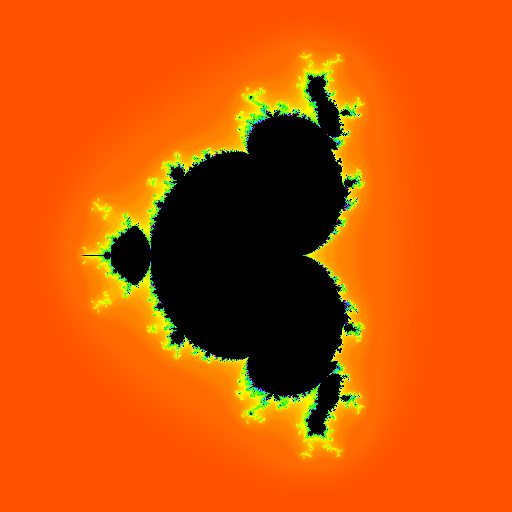}}\hfill
\subfloat{\includegraphics{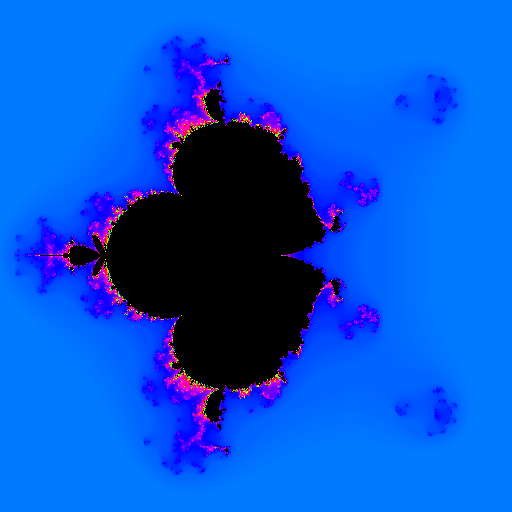}}\\
\subfloat{\includegraphics{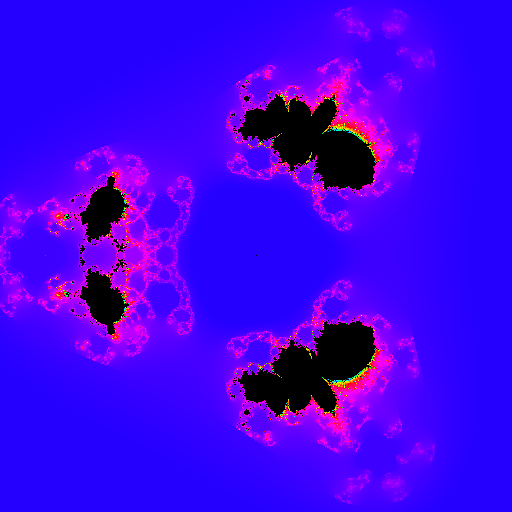}}\hfill
\subfloat{\includegraphics{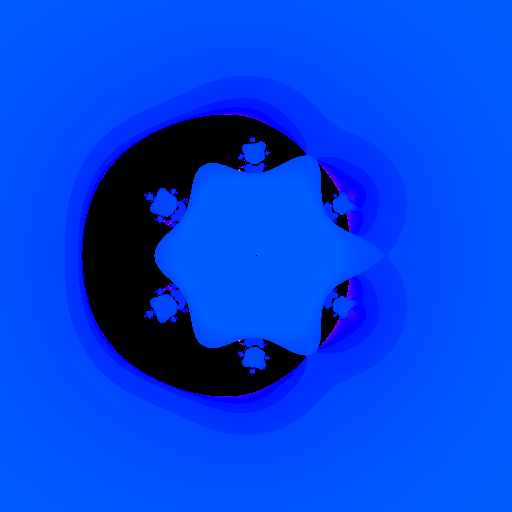}}\hfill
\subfloat{\includegraphics{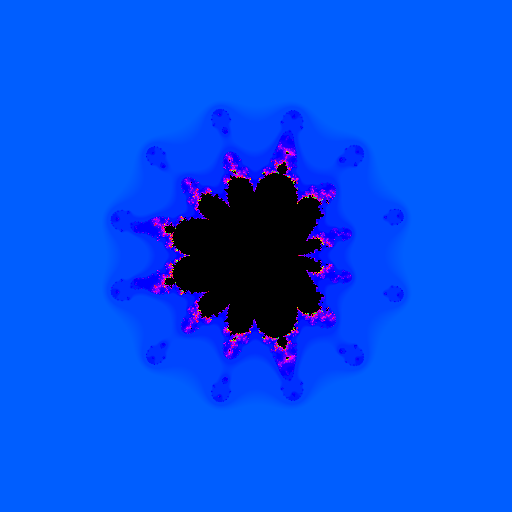}}\hfill
\subfloat{\includegraphics{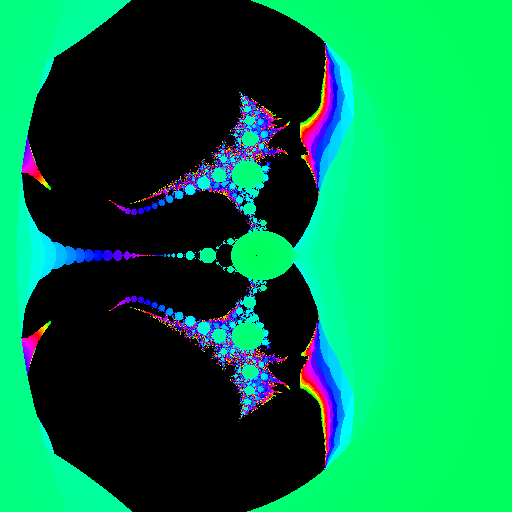}}
\caption{Goodware samples.}
\end{figure}

\begin{figure}[!ht]
\setkeys{Gin}{width=0.24\linewidth}
\subfloat{\includegraphics{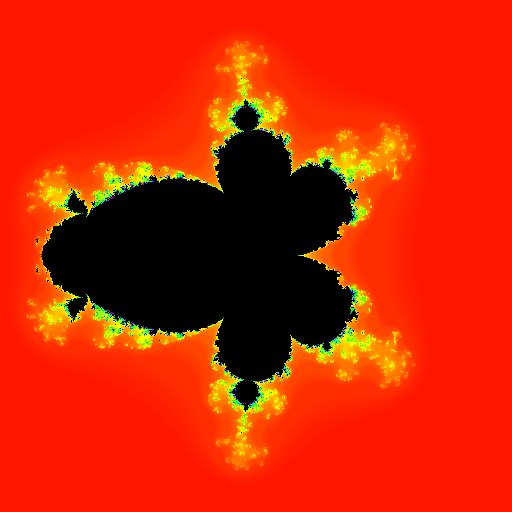}}\hfill
\subfloat{\includegraphics{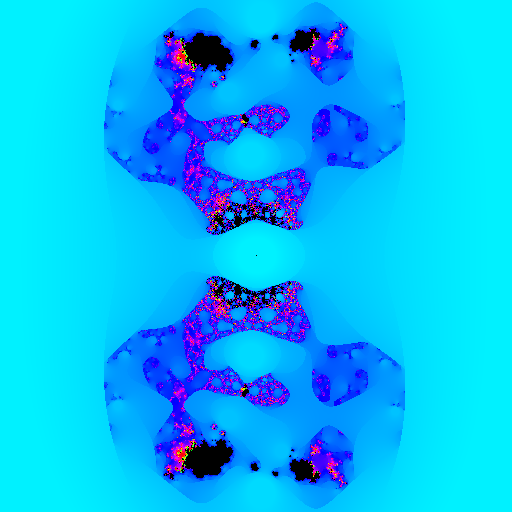}}\hfill
\subfloat{\includegraphics{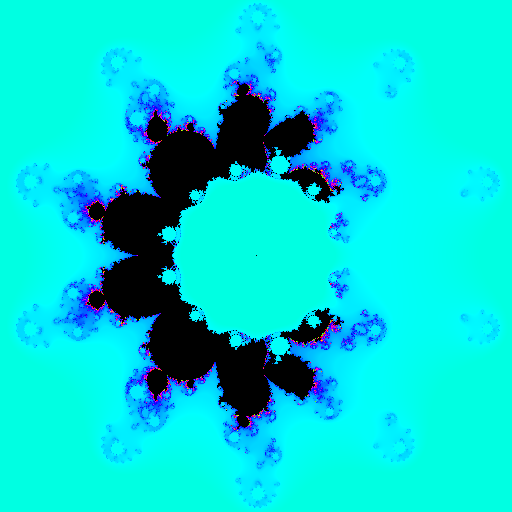}}\hfill
\subfloat{\includegraphics{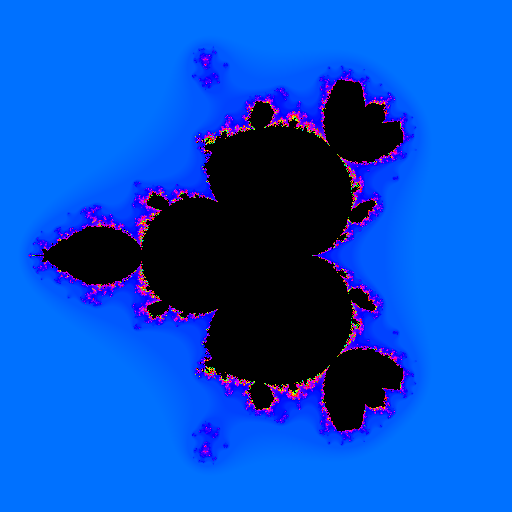}}\\
\subfloat{\includegraphics{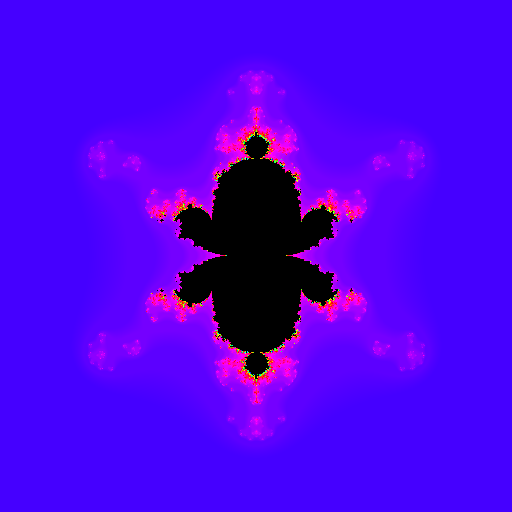}}\hfill
\subfloat{\includegraphics{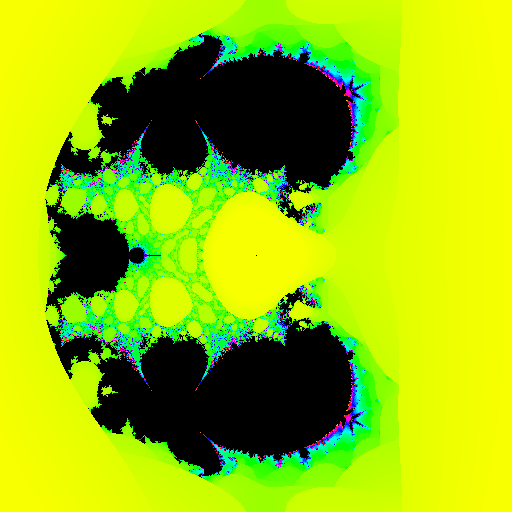}}\hfill
\subfloat{\includegraphics{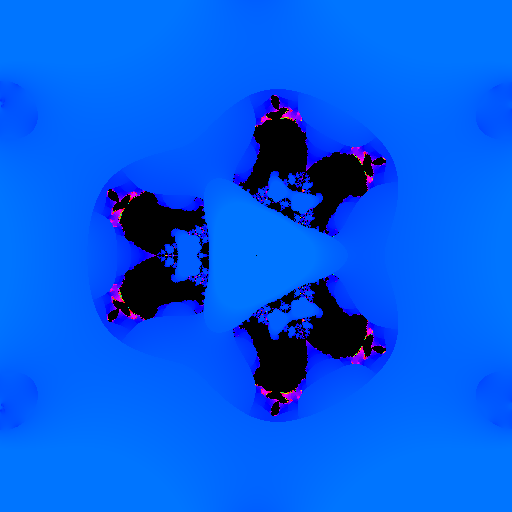}}\hfill
\subfloat{\includegraphics{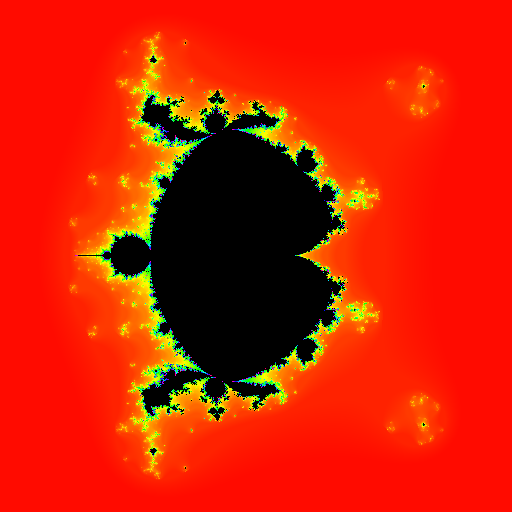}}
\caption{Goodware samples.}
\end{figure}

\begin{figure}[!ht]
\setkeys{Gin}{width=0.24\linewidth}
\subfloat{\includegraphics{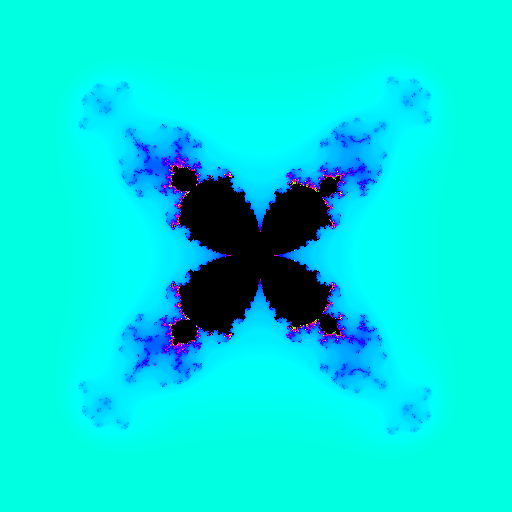}}\hfill
\subfloat{\includegraphics{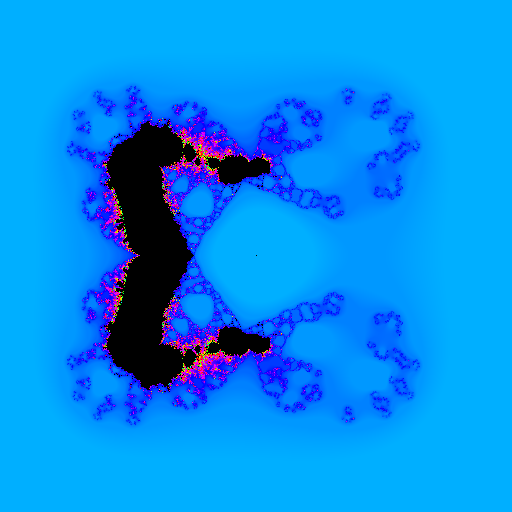}}\hfill
\subfloat{\includegraphics{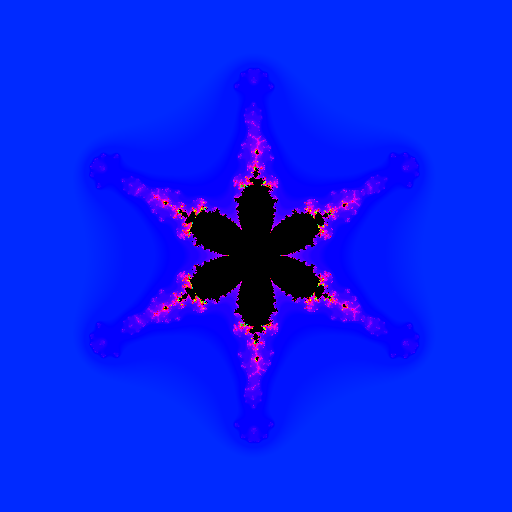}}\hfill
\subfloat{\includegraphics{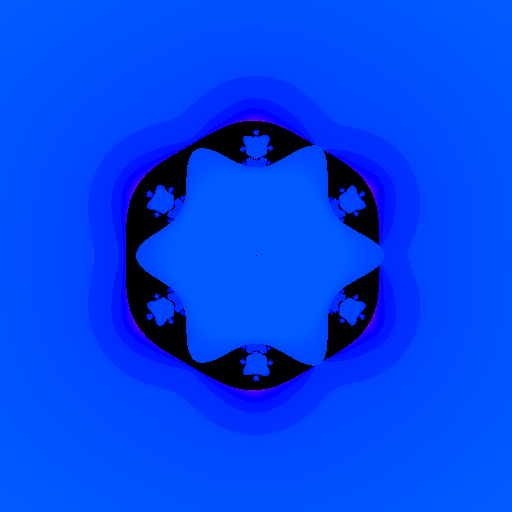}}\\
\subfloat{\includegraphics{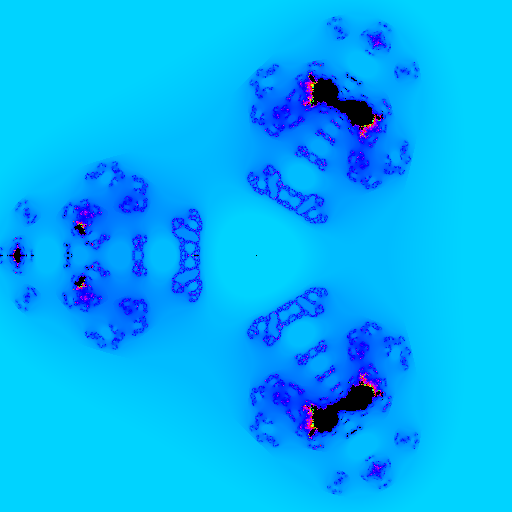}}\hfill
\subfloat{\includegraphics{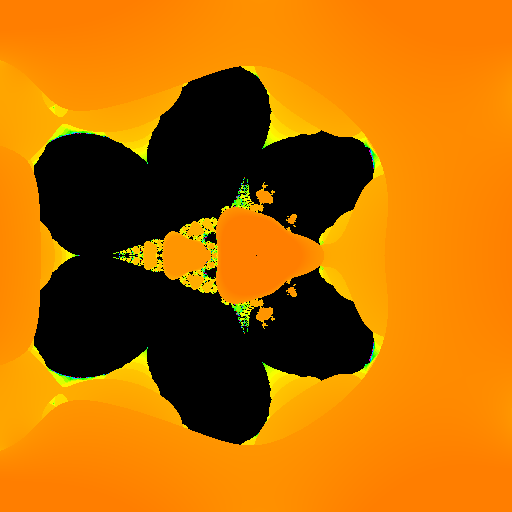}}\hfill
\subfloat{\includegraphics{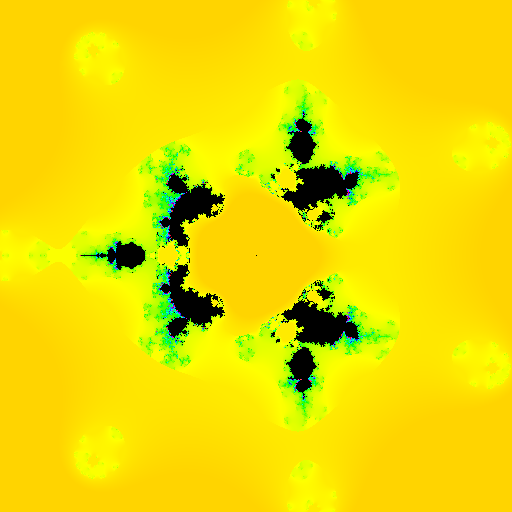}}\hfill
\subfloat{\includegraphics{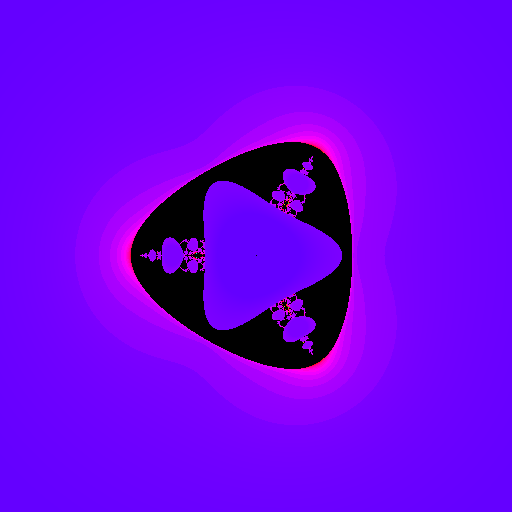}}
\caption{Goodware samples.}
\end{figure}

 \bibliographystyle{elsarticle-num} 
 \bibliography{cas-refs}





\end{document}